\documentclass[twocolumn]{aastex61}

\usepackage{threeparttable}

\newcommand*{\rom}[1]{\expandafter\@slowromancap\romannumeral #1@}

\newcommand{\msun}{M$_{\odot}$}
\newcommand{\myr}{M$_\odot$~yr$^{-1}$} 
\newcommand{\myrkpc}{M$_\odot$~yr$^{-1}$~kpc$^{-2}$} 

\newcommand{\ha}{H$\alpha$}
\newcommand{\hb}{H$\beta$}
\newcommand{\nii}{[N{\sc II}]}
\newcommand{\oiii}{[O{\sc III}]}
\newcommand{\sii}{[S{\sc II}]}

\newcommand{\loghn}{log(\nii/\ha)}
\newcommand{\logohb}{log(\oiii/\hb)}
\newcommand{\logsh}{log(\sii/\ha)}
\newcommand{\kms}{km\,s$^{-1}$}

\newcommand{\ergs}{erg s$^{-1}$}

\received{September 10, 2017}
\accepted{November 15, 2017}
\published{December 20, 2017}

\shorttitle{3C 298 Quasar Host Galaxy}
\shortauthors{Vayner et al.}

\begin{document}

\title{Galactic-Scale Feedback Observed in the 3C 298 Quasar Host Galaxy}

\correspondingauthor{Andrey Vayner}
\email{avayner@ucsd.edu}

\author[0000-0002-0710-3729]{Andrey Vayner}
\affiliation{Department of Physics, University of California San Diego, 
9500 Gilman Drive 
La Jolla, CA 92093 USA}
\affiliation{Center for Astrophysics \& Space Sciences, University of California San Diego, 9500 Gilman Drive La Jolla, CA 92093 USA}

\author[0000-0003-1034-8054]{Shelley A. Wright}
\affiliation{Department of Physics, University of California San Diego, 
9500 Gilman Drive 
La Jolla, CA 92093 USA}
\affiliation{Center for Astrophysics \& Space Sciences, University of California San Diego,
9500 Gilman Drive 
La Jolla, CA 92093 USA}

\author{Norman Murray}
\affiliation{Canadian Institute for Theoretical Astrophysics, University of Toronto, 60 St. George Street, Toronto, ON M5S 3H8, Canada}
\affiliation{Canada Research Chair in Theoretical Astrophysics}

\author[0000-0003-3498-2973]{Lee Armus}
\affiliation{Spitzer Science Center, California Institute of Technology, 1200 E. California Blvd., Pasadena, CA 91125 USA}

\author{James E. Larkin}
\affiliation{Department of Physics and Astronomy, University of California, Los Angeles, CA 90095 USA}

\author{Etsuko Mieda}
\affiliation{NRC Herzberg Astronomy and Astrophysics, 5071 West Saanich Rd, Victoria, BC, V9E 2E7, Canada}

\begin{abstract}
We present high angular resolution multi-wavelength data of the 3C 298 radio-loud quasar host galaxy (z=1.439) taken using the W.M. Keck Observatory OSIRIS integral field spectrograph (IFS) with adaptive optics, Atacama Large Millimeter/submillimeter Array (ALMA), Hubble Space Telescope (HST) WFC3, and the Very Large Array (VLA). Extended emission is detected in the rest-frame optical nebular emission lines \hb, \oiii, \ha, \nii, and \sii, as well as molecular lines CO (J=3-2) and (J=5-4). Along the path of 3C 298's relativistic jets we detect conical outflows in ionized gas emission with velocities up to 1700 \kms~and outflow rate of 450-1500 \myr~extended over 12 kpc. Near the spatial center of the conical outflow, CO (J=3-2) emission shows a molecular gas disc with a rotational velocity of $\pm$150\kms and total molecular mass ($\rm M_{H_{2}}$) of 6.6$\pm0.36\times10^{9}~$\msun. On the molecular disc's blueshifted side we observe broad extended emission due to a molecular outflow with a rate of 2300 \myr~and depletion time scale of 3 Myr. We detect no narrow \ha~emission in the outflow regions, suggesting a limit on star formation of 0.3 \myrkpc. Quasar driven winds are evacuating the molecular gas reservoir thereby directly impacting star formation in the host galaxy. The observed mass of the supermassive black hole is $10^{9.37-9.56}~$\msun~and we determine a dynamical bulge mass of M$\rm_{bulge}=$ 1-1.7$\rm \times10^{10}\frac{R}{1.6 kpc}~$\msun. The bulge mass of 3C 298 resides 2-2.5 orders of magnitude below the expected value from the local galactic bulge - supermassive black hole mass (M$\rm_{bulge}-M_{BH}$) relationship. A second galactic disc observed in nebular emission is offset from the quasar by 9 kpc suggesting the system is an intermediate stage merger. These results show that galactic scale negative feedback is occurring early in the merger phase of 3C 298, well before the coalescence of the galactic nuclei and assembly on the local M$\rm_{bulge}-M_{BH}$ relationship.
\end{abstract}



\section{Introduction} \label{sec:intro}
Quasars are the most luminous active galactic nuclei (AGN), whose supermassive black holes (SMBHs) are often fueled by large galaxy mergers \citep{Treister12}. AGN accretion discs are thought to drive energetic winds \citep{Murray95} and/or relativistic jets that suppress star formation \citep{Hopkins12,Zubovas14,Costa15}, thereby impacting galactic structure and evolution. This is one of the leading theoretical \citep{DiMatteo05,Faucher12,Barai17,Alcazar17} pictures used to explain correlations between SMBH masses and galactic stellar masses \citep{Magorrian98,Gebhardt00,Ferrarese00,Kormendy13} and luminosity function of local massive galaxies \citep{Benson03}. The sphere of influence of SMBHs, inside of which their gravity dominates that of the stars, gas, and dark matter, range from a few 10s to 100 pc, while the local scaling relations of M$\rm_{bulge}-M_{BH}$ apply on stellar bulge scales, i.e., several kpc. The SMBH energy output can be orders of magnitude higher than the binding energy of galactic bulges; therefore energy injected by active SMBHs into the interstellar medium (ISM) could be efficient at impacting the stellar mass history of its host galaxy. Since the bulk of stellar mass is formed at high-redshift (z$>1$) \citep{Gallazzi08}, it is critical to study the effects of AGN activity during the assembly periods of their host galaxies at high-redshift. 

Observational studies have found that nearby quasars and ULIRGs (Ultra-Luminous Infrared Galaxies) show evidence of large scale ionized \citep{Greene12,Rupke11,LiuG13,Harrison14} and molecular outflows (e.g., \citealt{Cicone14,Sun14,Stone16,Veilleux17}) allowing for detailed studies of feedback physics. However the computed outflow rates in nearby systems are not sufficient to impact the stellar mass assembly history since these galaxies have already done the bulk of their growth. In contrast, there is minimal observational evidence for AGN activity directly affecting star formation at the peak epoch (1$<$z$<$3) of galaxy formation and black hole accretion \citep{Delvecchio14,Madau14}. Recent studies of distant quasars and radio-loud galaxies have shown evidence of large scale (5-20 kpc) outflows driven by AGN activity that theoretically should be powerful enough to quench star formation  \citep{Nesvadba08,Steinbring11,Cano-Diaz12,Harrison12,Brusa15,Carniani15}. Although it has still remained challenging to directly associate large scale outflows with the suppression of star formation of AGN host galaxies \citep{Cano-Diaz12,Cresci15,Carniani16}. These studies have provided intriguing clues as to the nature of quasar feedback. Yet compared to nearby quasars there is little known about z $>$ 1 host galaxies (i.e., stellar mass, dynamics, and morphologies) and the true effects of feedback on host galaxy star formation. A key missing result is one that connects quasar and jet driven outflows with the observed galactic scale winds, star formation activity, and molecular gas properties.  

Quasars outshine their host galaxies by an order of magnitude or more, making it observationally challenging to detect and study their faint underlying galaxies \citep{Lehnert99,Hutchings02,Jahnke04,Falomo05,Floyd13,Glikman15}. The sizes of distant (z $>$ 1; look-back time of 9.25 Gyr) galaxies are small ($\sim$ 1\arcsec), roughly the same angular size as ground-based ``seeing" and space-based instrument resolution and contrast sensitivity. The combination of near-infrared integral field spectroscopy (IFS) with laser-guide star adaptive optics (LGS-AO) allows for the effective separation of quasar emission from the host galaxy. This is achieved by using a pristine point spread function (PSF) generated by the quasar broad-line and continuum emission from the IFS data cube \citep{Inskip11,Vayner16}. 

We have started the QUART (Quasar hosts Unveiled by high Angular Resolution Techniques) survey of high-redshift (1.3$<$z$<$2.6) quasars using W.M Keck Observatory laser guide star adaptive optics (LGS-AO) observations to resolve and study quasar host galaxy properties with a rich multi-wavelength data sets. Herein we present the first paper of this survey on the individual z=1.439 (look-back time 9.3 Gyr) radio loud quasar 3C 298. We combine high spatial resolution observations to reveal the complex morphology, dynamics, and energetics of the 3C 298 host galaxy. Using OSIRIS and Keck LGS-AO, we are able to map the kinematics and intensity of galactic nebular emission lines \hb, \oiii, \ha, \nii and \sii~at$\sim$1.4 kiloparsec (kpc) resolution. We present the multi-wavelength data sets in \S \ref{sec:obs} and describe the data reduction and analysis techniques for each instrument in \S \ref{sec:data}. We discuss the dynamics and energetics of ionized and molecular gas in \S \ref{sec:outflow}, and in \S \ref{sec:discussion} we discuss our results and overall interpretation. We suggest the reader refers to Figure \ref{fig:cartoon} while reading the manuscript, which summarizes the observed structure and properties of the 3C 298 host galaxy. Throughout the paper we assume a $\Lambda$-dominated cosmology \citep{Planck13} with $\Omega_{M}$=0.308, $\Omega_{\Lambda}$=0.692, and H$_{o}$=67.8 \kms~ Mpc$^{-1}$

\section{Observations}\label{sec:obs}

We present new observations of 3C 298 using Keck OSIRIS AO and ALMA band 4 and 6. These observations are coupled with archival Hubble Space Telescope (HST) WFC3 and Very Large Array (VLA) imaging data. 

\subsection{Keck: OSIRIS}
Observations were taken using the integral field spectrograph OSIRIS \citep{Larkin06} with the upgraded grating \citep{Mieda14} behind the laser guide star adaptive optics (LGS-AO) system at W.M. Keck Observatory on May 19 and 20, 2014 (UT). The quasar was used for tip/tilt correction while the laser tuned to 589.2 nm created an artificial star on-axis for higher order corrections. We used the Hn3 (May 19) and Jn1 (May 20) filters with a plate scale of 100 milli-arcseconds (mas) per lenslet with a position-angle of 103$^\circ$. In these modes OSIRIS has a field of view of 3.2\arcsec$\times$6.4\arcsec in Jn1 and 4.8\arcsec$\times$6.4\arcsec in Hn3. We took four 600s exposures on-source in each filter, plus an additional 600s pure-sky frame. Each night immediately after the quasar observations we observed the standard star HD136754 for telluric and flux calibrations. Both nights were photometric with near-infrared seeing of 0.4-0.5\arcsec. 

\subsection{ALMA: Band 4 and 6}

Early ALMA science (cycle 2, 3) band 4 and 6 observations (2013.1.01359.S, 2015.1.01090.S, PI: Vayner) were aimed at observing the rotational molecular transition of CO J=3-2 and J=5-4 in emission to map the distribution and kinematics of the molecular gas in 3C 298. One 1.8745 GHz spectral window was centered on CO (J=3-2) (141.87 GHz) and CO (J=5-4) (236.43 GHz) while three additional spectral windows set up to map the continuum in each band. The effective velocity bandwidth per spectral window was approximately 4,000 \kms~for band 4 and 2,400 \kms~for band 6. Observations were taken in an extended configuration with an approximate angular resolution of $\sim$0.4\arcsec~and $\sim0.3$\arcsec~for band 4 and 6, respectively. In Table \ref{ALMAtable:summary} we summarize the observational setup for each band.

\begin{deluxetable*}{llcllclcc}
\tablecaption{ALMA Cycle 2 and 3 observations summary\label{ALMAtable:summary}}
\tablehead{\colhead{Date} & \colhead{Band} & \colhead{Central frequencies} & \colhead{Integration} & \colhead{PWV} & \colhead{Antennae} & \colhead{Beam} & \colhead{Line\tablenotemark{a}} &\colhead{Continuum}\\\colhead{} & \colhead{} & \colhead{(GHz)} & \colhead{Time (min)} & \colhead{(mm)} & \colhead{$\#$}& \colhead{Size} & \colhead{$\sigma$/beam (mJy)} & \colhead{$\sigma$/beam ($\mu$Jy)}}
\startdata
2015 Aug 6 & 4 & 141.86 & 27.84 & 4.5 & 39 &0.44\arcsec$\times$0.41\arcsec&0.5&48\\
2016 Sep 9 & 4 & 141.85, 140.00, 129.97, 128.01 & 24.19 & 2.5 & 37& 0.39\arcsec$\times$0.30\arcsec&0.22&46\\
2016 Sep 17 & 6 & 236.42, 234.5, 220.99, 219.11 & 19.65 & 0.7 & 38& 0.28\arcsec$\times$0.18\arcsec&0.37&44\\
\enddata
\tablenotetext{a}{Computed per 34\kms~channels}
\end{deluxetable*}

\section{Data reduction and Analysis \label{sec:data}}

\subsection{OSIRIS: Data reduction \label{subsec:osiris}}
The data was reduced with the OSIRIS data reduction pipeline version 3.2, which performs standard near-infrared IFS reduction procedures: dark subtraction; adjust channel levels; remove crosstalk; glitch identification; clean cosmic rays; extract spectra; assemble data cube; and correct dispersion. We used the \textit{scaled sky subtraction} routine for the Jn1 data cubes, which uses families of atmospheric OH-emission lines between multiple frames for a cleaner subtraction. For Hn3 data we used our own custom sky subtraction routine that scales only the nearest three OH lines in proximity to the quasar \ha~emission line. This produced better residuals compared to the pipeline's scaled sky routine in Hn3. Inspection of individual spaxels in the scaled sky subtracted data cubes still revealed strong OH-sky line residuals, typically over 2-3 spectral pixels in the wings of the \ha~emission line. These spectral pixels were linearly interpolated using the slope from the neighboring two pixels around the strong residuals. The telluric spectrum of the calibration star was extracted over the seeing halo and had its continuum divided by a 9400K blackbody function with its hydrogen absorption lines removed. The 1D telluric spectrum was normalized and divided into the quasar data cubes. Individual data cubes were then shifted to a common position and combined using a $3\sigma$ clipping algorithm, which is part of the OSIRIS data reduction pipeline. Finally we applied flux calibration to all quasar data cube by using the telluric corrected standard star spectrum and scaling the DN/s/channel to match the expected Jn1 and Hn3 band flux (erg/s/cm$^{2}$). 
 
\subsection{OSIRIS: PSF construction and subtraction}
Both the quasar broad line emission (\hb~and \ha) and quasar continuum originate from gas on parsec scales from the SMBH, and are therefore spatially unresolved in our OSIRIS observations. Wavelength channels of the broad line emission and/or continuum can be used to construct a pristine quasar image that can then be used for PSF subtraction in the reduced data cube. In \citet{Vayner16} we discuss in greater detail our PSF subtraction routine and its performance on a set of luminous type-1 radio quiet quasars at z$\sim$2. In brief, we used broad emission line/continuum channels that should not overlap with the host galaxy emission with spectral channels offset by 5,000-10,000 \kms~from the quasar redshift. We carefully select spectral channels that do not coincide with OH-emission lines or regions of low transparency in the near-infrared. PSFs are generated by combining individual data channels and are scaled to the peak pixel values. This empirical PSF is then subtracted from the entire data cube while re-scaling to the peak pixel value of the quasar emission per wavelength channel.

After data reduction, the 100 mas mode in OSIRIS suffers from flux mis-assignment between adjacent spaxels given its enlarged pupil size in the instrument. The reductions still conserve the integrated flux of the source, but neighboring pixels can receive a $\sim$10\% mis-assignment of flux. This effect is seen in bright stars (H $<$ 16 mag) with excellent AO correction, where flux is mis-assigned from the bright central spaxel to the row above and below the centroid position of the point source. This effect can be easily identified by evidence of spaxels with an inaccurate spectral shape. Our PSF subtraction routine removes a significant portion of flux from the PSF in these spaxels but they generally have stronger post-PSF subtraction residuals. We masked the affected spaxles with background flux values calculated by taking a standard deviation in a 1\arcsec$\times$1\arcsec sky region. We fit a 2D Gaussian to the PSF image and measure a full-width half-maximum (FWHM) of 0.127\arcsec and 0.113\arcsec in Hn3 and Jn1, respectively. After PSF subtraction of the quasar data cubes, we smooth the Hn3 and Jn1 data sets to a common resolution with a beam FWHM of 0.2\arcsec~to improve the signal-to-noise (SNR) in the diffuse parts of the host galaxy.

\subsection{OSIRIS: Kinematics}\label{sec:OSIRIS_kinematics}

In this section we investigate the kinematics properties of nebular emission lines in the host galaxy of 3C 298. We inspect $\textit{all}$ individual spectra that overlap between the two observing modes (Jn1 and Hn3), which amounts to approximately 2,640 spectra. An emission line is identified to be real if the peak intensity is at least 3$\times$ greater than the noise per wavelength channel, and the emission line dispersion is larger than the instrumental resolution (0.206 nm for Hn3 and 0.174 nm for Jn1). 

A single Gaussian profile provides a good fit to the 486.1 nm \hb~emission line that is redshifted into Jn1 at an observed wavelength of 1186 nm. The 495.9, 500.7 nm \oiii~lines are redshifted into Jn1 band at an observed wavelength of 1210 and 1221 nm. The 495.9, 500.7 nm \oiii~lines are fit simultaneously, each with a single Gaussian profile. The position and width of the 495.9 nm line are held fixed to the redshift and width of 500.7 nm line at each spaxel. The line ratio between the \oiii~lines are held fixed at 1:2.98 \citep{Storey99}. The 495.9, 500.7 \oiii~lines in several spaxels in the north-west and south-east regions require two Gaussian profiles for a good fit. The \nii~654.9, 658.5 nm, \ha~656.3 nm, and \sii~671.7 673.1 nm are redshifted into the Hn3 filter. Spaxels with detected \ha~and both \nii~emission lines are fit together with three Gaussian profiles. The position and width of the \nii~lines are fixed to the \ha~redshift and width, with a flux ratio between \nii~654.9, 658.5 nm of 1:2.95. Generally a single Gaussian profile fit to \ha~and each \nii~line provide a satisfactory fit ($\chi^{2}_{R}\sim1-2$), with the exception of several spaxels in the south-east region, where an additional broad \ha~component is necessary. The \ha, \hb~and \oiii~broad-line components have similar velocity dispersions and offsets. The lower SNR of the \sii~doublet make it challenging to fit a single Gaussian profile to each emission line, therefore each spaxel where \sii~is detected we fit a single Gaussian profile to the combined signal. Regions where no \nii~or \sii~lines are detected have only a single Gaussian fit to \ha. Limits on \nii~and \sii~lines are derived by inserting a Gaussian profile with the same width as \ha~at the expected location of the emission line based on \ha's redshift, with a peak flux 2$\times$ greater than the standard deviation of the noise. Similarly to a real detection we integrate the inserted emission line to obtain a limit. 

To construct 2D flux maps we integrate each emission line from $-3\sigma$ to $+3\sigma$, where $\sigma$ is derived from the line fit. Error on the line flux is calculated by taking a standard deviation of every 3 spectral channels and summing in quadrature over the same spectral region where the nebular emission line is integrated. Velocity maps for each emission line are relative to the redshift of the quasar's broad line region, which is calculated by fitting a single Gaussian profile to the broad \hb~line constructed by spatially integrating the cube over the seeing halo before PSF subtraction. The velocity dispersion map has the instrumental PSF subtracted out in quadrature using the width of OH emission sky lines. Errors on velocity offsets and dispersions are based on 1$\sigma$ errors associated with the least squares fit. 

\begin{figure*}[!th]
    \centering
    \includegraphics[width=6.2in]{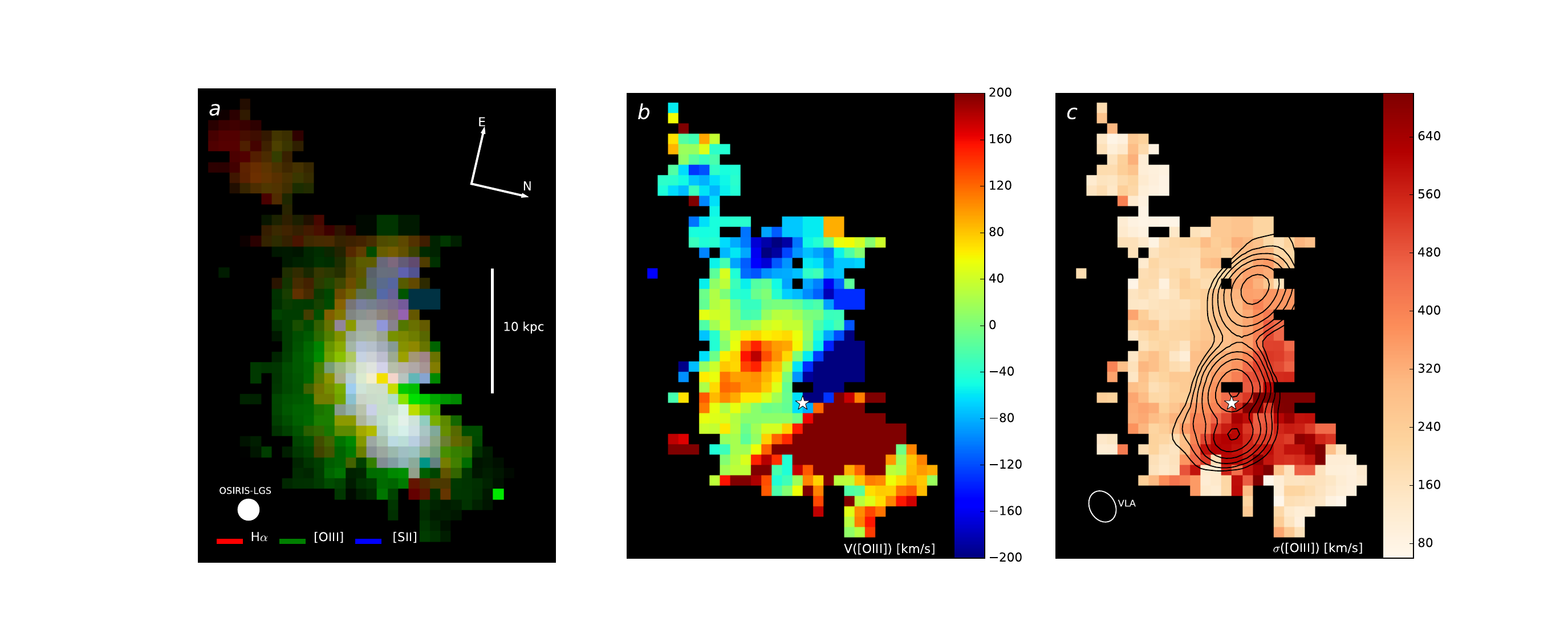}
    \caption{OSIRIS observations of 3C 298 nebular emission line intensities and \oiii~kinematic maps. (a) Three colour intensity map of nebular emission lines: \ha (red), \oiii (green), and \sii (blue). FWHM of the OSIRIS beam is shown as a white circle on the bottom left. (b) Radial velocity offset (\kms) of the \oiii~line relative to the redshift of the quasar. The white star shows the location of the subtracted luminous quasar. (c) Velocity dispersion (\kms) map of \oiii emission. The black contours are 8485.100 MHz VLA observations of radio synchrotron emission from the quasar jet and lobes. The largest velocity offset and dispersion corresponds to the location of jet and lobes. This spatial correspondence provides evidence of a quasar driven outflow in the ionized ISM with an ionized outflow rate of 450-1500 \myr.}
    \label{fig:O3_maps}
\end{figure*}

In Figure \ref{fig:O3_maps} we show the line integrated emission of \ha, \oiii~and \sii~in a three colour image composite alongside the \oiii~radial velocity and dispersion maps of 3C 298. A distinct extended broad velocity emission region is co-spatial with radio synchrotron emission emanating from extended jet/lobes. The radio VLA observations are taken from \cite{mantovani13} (Project code: AJ206). We downloaded fully reduced clean map of 3C 298 at 8485.100 MHz \footnote{http://db.ira.inaf.it/aj206-fm/}. The centroid of the point source with the flattest spectral slope is associated with the optical location of the quasar. We extract the image centered on the quasar and rotate to a position angle of 103$^{\circ}$ to match our OSIRIS observations.

\begin{figure*}[!th]
   \centering
    \includegraphics[width=6in]{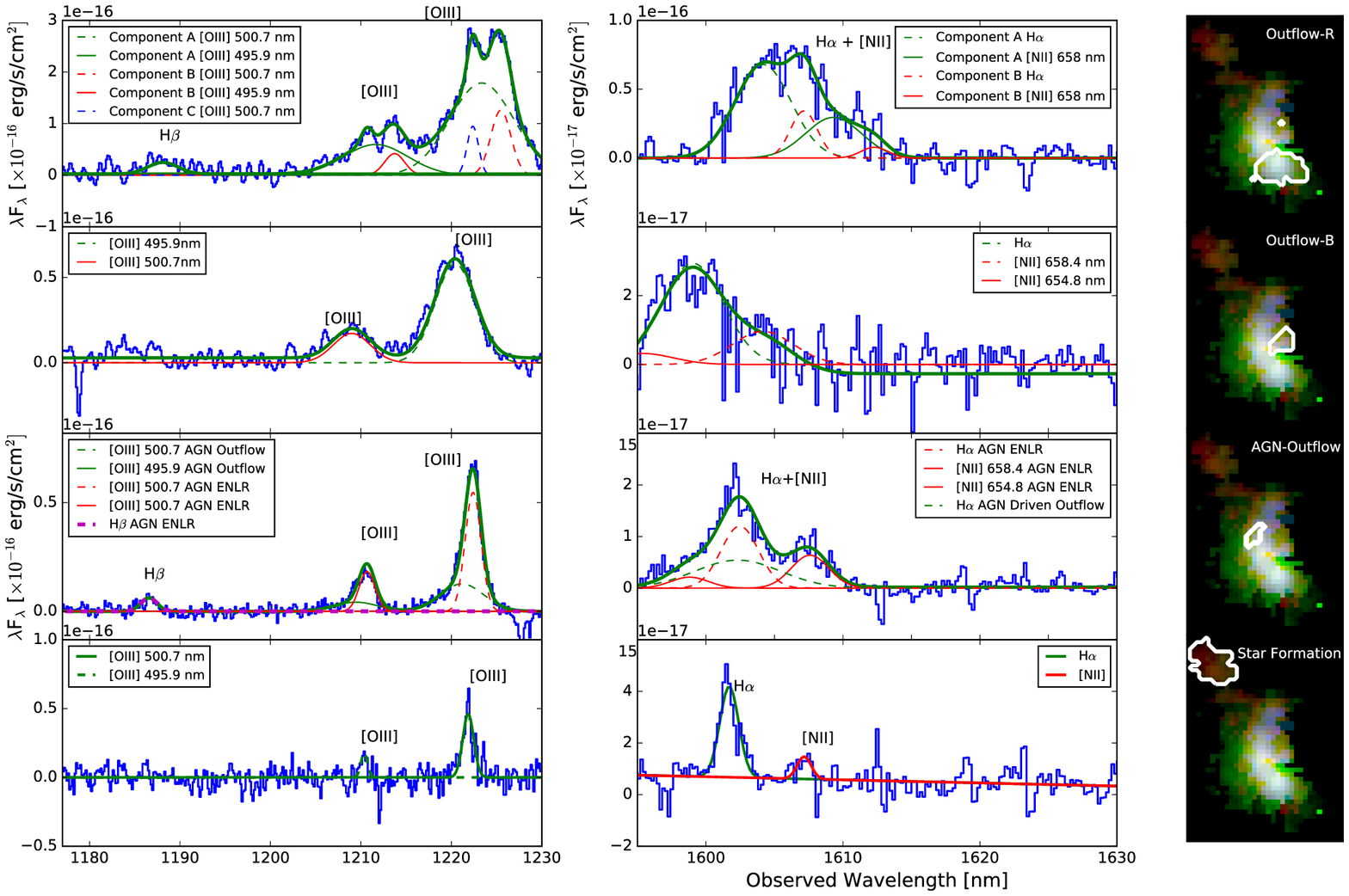}
    \caption{OSIRIS near-infrared spectra for distinct photoionized regions in the host galaxy of 3C 298. J and H band spectra are shown in the left and middle panels, respectively, with observed wavelengths (nm) and calibrated flux (F$_{\lambda}$). The three-colour composite image (right) shows the regions over which the spectra were extracted. Multiple Gaussian functions are fit to each of the emission lines: \hb, [OIII], \ha, and  [NII] in all the spectra. The green curves show the combined fit to each spectrum, and each dotted line represents an individual component of the fit. The redshifted outflow region is best fit with a double Gaussian component (labeled A and B in top panel) to each nebular emission line, that suggests conical expansion of the gas along the line-of-sight. The blueshifted outflow region is best modeled with a single Gaussian component to each emission line. The presence of broad, blueshifted emission lines in \oiii~and~\ha~(labeled `AGN Outflow', third panel from top), indicating outflowing gas, roughly coincident with the dynamical center of the secondary galaxy in the merger. This outflow region is not associated with strong star formation, based on line ratio diagnostics in Figure \ref{fig:n2ha_BPT}. The star forming region is best fit with a single narrow Gaussian component to each emission line.}
    \label{fig:all_spec}
\end{figure*}

We generate integrated spectra for three distinct outflow regions identified based on their radial velocity ($\pm$400 km/s) and dispersion (V$_{\sigma}>$500\kms), see Figure \ref{fig:O3_maps}. For the purposes of the figure, the north-western outflow will be identified as the redshifted outflow or ``Outflow-R", and the north-eastern outflow will be identified as the blueshifted outflow or ``Outflow-B", and the third outflow in the south-eastern direction will be identified as ``AGN-Outflow". We believe AGN-outflow belongs to a secondary nucleus in the 3C 298 system due to its isolated nature, and since the ionized emission does not extend from the quasar and does not coincide with the quasar jet/lobes. The AGN-outflow is in close proximity to the dynamical center of a rotating disc in the 3C 298 system that belongs to a second merging galaxy (see section \ref{sec:vel_model} for further discussion). Each spectra are then fit with multiple Gaussian profiles. \textit{Outflow-R} is best fit with a combination of two relatively broad (V$_{\sigma}\sim$500 km/s) Gaussian profiles in both \oiii~lines, \ha, and \nii, which we interpret as a signature of outflowing gas along the line of sight. The 500.7 nm \oiii~line requires an additional relatively narrow (V$_{\sigma}\sim$ 80 km/s) component for a good fit that has no counterpart in \ha, \nii, or \sii. A faint \hb~line is detected, and is fit with a single Gaussian component that potentially matches the broad components of \oiii~and \ha. See the top row of Figure \ref{fig:all_spec} for the integrated spectrum in J and H band along with the fit. Components A (green curve) and B (red curve) are the broad lines, while component C (blue curve) is the narrow line found only in \oiii. The white contour in the right column shows the region over which the data cube is spatially integrated. The spectrum of the blueshifted outflow (\textit{Outflow-B}) region is best fit with a single broad Gaussian component in \hb, \oiii, \ha~and \nii; second row of Figure \ref{fig:all_spec} shows the spectra along with the fit. In the redshifted outflow region the lines are very broad, so the \sii~doublet blends together, making it very hard to fit the individual lines. The lines are slightly narrower in the blueshifted outflow region, and each emission line in \oiii~and \ha~require only a single Gaussian component. This made it possible to fit the \sii~doublet using a single Gaussian profile for each emission line. 

We construct a high SNR spectrum over the entire \textit{AGN-Outflow} region. \ha~and \oiii~are fit with two components: a broad-line blueshifted Gaussian for the outflow and a narrow component that we interpret as part of the AGN/quasar narrow-line region. \hb, \nii, and \sii~are fit with a single Guassian narrow-line component.

\subsection{OSIRIS: Nebular Emission line Diagnostics}\label{sec:OSIRIS_BPT}

In this section we investigate potential ionizing sources for distinct regions of the host galaxy. Line ratio maps are constructed by taking ratios of integrated flux maps \oiii, \ha, \nii, and \sii. The \hb~line is only detected in a small number of individual spaxels; sensitivity and dust obscuration prohibits us from detecting this line over similar sized regions to other detected nebular emission lines. We construct an \hb~map over the region where \ha~is detected by assuming case B recombination ($F_{H\beta}=F_{H\alpha}/2.89$).

\begin{figure*}[!th]
    \centering
    \includegraphics[width=6.5in]{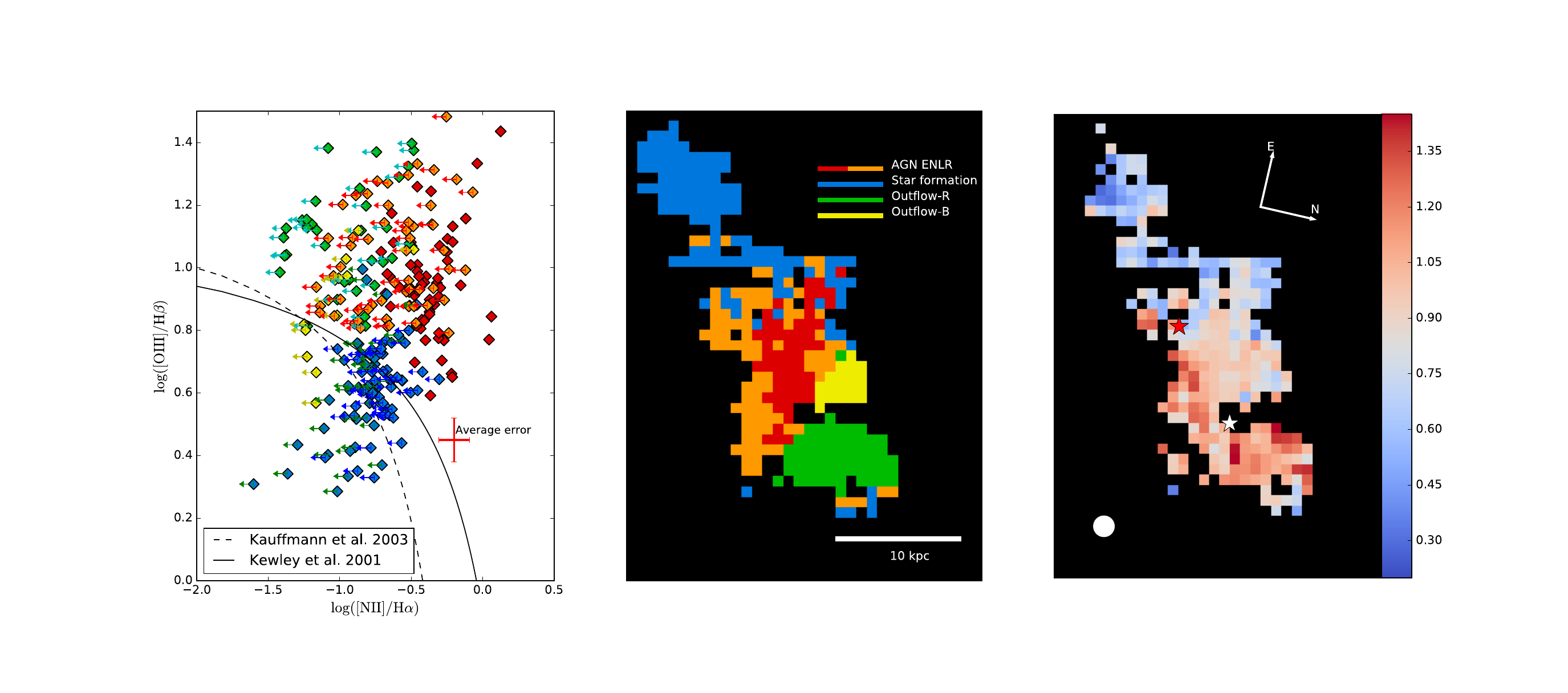}
    \caption{Photoionization diagnostics of ionized gas in the 3C 298 system. (LEFT) The \logohb~vs \loghn~nebular line ratio diagram of the 3C 298 host galaxy with empirical \citep{Kauffmann03} and theoretical \citep{Kewley01} curves that delineate the separation between star formation and AGN. Nebular line ratios that lie above and to the right of the curves are considered to be photoionzed by a hard radiation source such as a quasar/AGN or shocks, and values below the curves are considered to be ionized by O-stars in HII regions. (MIDDLE) Galaxy map coloured to match the line ratio diagnostic diagram.  Red and orange colours represent regions photoionization by the quasar and secondary AGN in an extended narrow line region (ENLR). Regions with upper limits on \loghn~are labelled orange, while red regions have directly measured 658.4nm \nii line. The blue region shows gas whose line ratios are consistent with ionization by young massive stars. The green region is the redshifted outflow, and the yellow region is the blueshifted outflow. The redshifted outflow (green) shows a blend of photoionization from the quasar and radiative shocks due to interactions of the ISM with large scale outflows. This is further confirmed with the \logohb~vs. \logsh~diagnostic diagram as seen in Figure \ref{fig:osiris_bpt}. (RIGHT) The \logohb~ratio map illustrating different photoionization levels across the galaxy. The colours clearly cluster in distinct regions in the resolved host galaxy. The beam size is given by the solid white circle. The white star represents the location of the quasar and the red star and circle represents the location of the dynamical center of the secondary galaxy, where a potential secondary obscured, AGN exists.}
    \label{fig:n2ha_BPT}
\end{figure*}

We create three line ratio maps \logohb, \loghn, and \logsh. In Figure \ref{fig:n2ha_BPT} we plot the line ratios on a standard BPT diagram, \logohb~vs \loghn. Empirical \citep{Kauffmann03} and theoretical \citep{Kewley01} curves separating photoionization by O-stars versus quasar ionization are shown. Generally, values that lie above these curves represent photo-ionization by an AGN, while points below represent photoionization by newly formed O-type stars, tracing regions of active/recent star formation. 
Points on the BPT diagram are colour coded to match the line ratio map of \logohb~(middle Figure \ref{fig:n2ha_BPT}). The 1\arcsec~central region (red and orange in Figure \ref{fig:n2ha_BPT}) is mainly photoionized by the quasar in an extended narrow line region (ENLR), as shown by very high \logohb $\gtrsim$ 0.8 values. 658.5\nii~is detected over the red region in individual spaxels while over the orange region we place a 2$\sigma$ flux limit. Even within the limits the orange region fall in the AGN photoionzation portion of the BPT diagram due to very high \logohb~values. The star formation region (blue) has low \logohb$<$0.8 and \loghn$<$0.5 values, corresponding to ionization from newly formed stars. We generate a spectrum over the \textit{Star Formation} region and easily identify and fit a single Gaussian to \oiii, \ha, \nii~and a \hb~line with a SNR of 2 identified (bottom panel of Figure \ref{fig:all_spec}).

\begin{figure}
    \center
    \includegraphics[width=3.4in]{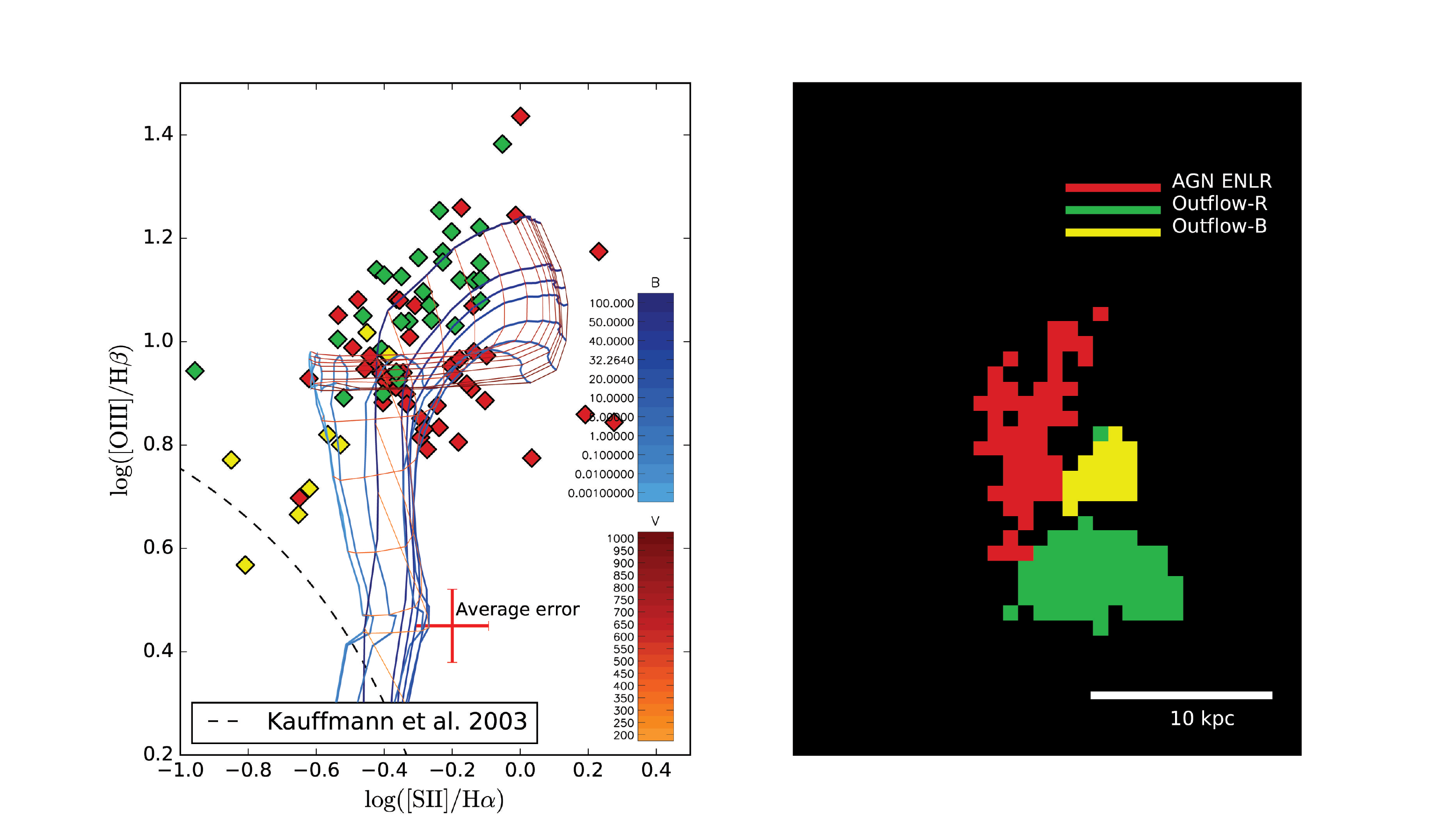}
    \caption{Photoionzation diagnostics of ionized gas in the 3C 298 system. (LEFT) We show diagram of \logohb~vs \logsh~with empirical dash curve\citep{Kauffmann03} that separates photoionization of star forming regions and AGN. The theoretical position of shocked ionization is overlaid on the BPT diagram \citep{Allen08}. Dark red/blue lines in the model represent higher velocity gas and stronger magnetic parameter. (RIGHT) Coloured values match the spatial regions of the host galaxy, where the green and yellow points are from the redshifted and blueshifted outflow regions. Ionization in these outflow regions are due to shocks moving at speeds of $\sim$1000\kms. The red region is consistent with values primarily ionized by AGN.}
    \label{fig:osiris_bpt}
\end{figure}

Star forming regions are spatially offset from the large outflows, and their kinematics and morphologies agree with a potential tidal feature induced by the merging galaxies. Outflow regions (yellow and green) correspond to the highest \logohb~line ratios. These regions are inconsistent with ionization by young stars, and are photoionized by quasars and shocks. Gas in outflow regions is moving at sufficiently high velocity so the Mach number of the wind is high and strong radiative shocks are expected. Using shock models from \citep{Allen08} we find that a large portion of observed line ratios agree with shock models for gas with electron density between $10^{2}-10^{3}$cm$^{-3}$ and shock velocities of 1000 km/s. In Figure \ref{fig:osiris_bpt} we plot the \logohb~vs \logsh~values over regions where \sii~was detected in individual spaxels. We overlay shock models on this BPT diagram, a large fraction of the nebular line ratios in the redshifted outflow region (green) and blueshifted outflow region (yellow) are consistent with these models. Dark red/blue lines in the model represent higher velocity gas and stronger magnetic parameter (see \cite{Allen08} for further details). Furthermore the region photoionized by AGN shows a spatial profile that drops off as $\rm 1/R^{2}$ in nebular emission, while in the region where we see powerful outflows the profile is more complex suggesting multiple ionizing sources. Both regions show similar extinction values as we discuss below.

In order to investigate how dust can alter our photoioniziation measurements, we construct a high SNR spectrum over the ENLR (red in Figure \ref{fig:n2ha_BPT}) to estimate the amount of dust obscuration from the \ha/\hb~ratio. Using the \citet{Calzetti00} law, we find an average ${\rm A_v }= 0.07$ over the ENLR (the red region in Figure-\ref{fig:n2ha_BPT}), based on an \ha/\hb~ratio of 2.9. This would on average increase the ratio of \logohb~only by a factor of 1.05 for the ENLR; this is not large enough to alter our results. We find that in portions of the the ENLR and redshifted outflow region the A$_{\rm v}$ value is consistent with zero extinction in the ionized gas. Similar results are found over the outflow regions suggesting negligible extinction in the ionized gas. Over the star forming region (blue region in Figure-\ref{fig:n2ha_BPT}) we measure an ${\rm A_v}$ value of 1.2 based on \ha/\hb~ratio of 4.0. The observed non-detection of \hb~in individual spaxels is consistent with the derived integrated extinction value and given the expected \hb~SNR relative to \oiii~and \ha~emission lines.

\subsection{OSIRIS: Velocity field modeling}\label{sec:vel_model}

To the south-east of the quasar there is a kinematic feature resembling a rotating galactic disc that is offset by $\sim200$ \kms~from the quasar BLR, with a projected rotational velocity of $\pm$170 \kms. In this section we outline the modeling done on the radial velocity map of \oiii~ emission to confirm the rotating disc nature of this feature. We isolate this feature for the disc fitting, using a box size of 0.7\arcsec$\times$1.7\arcsec centered on the gradient feature in the \oiii~radial velocity map. The boxed region is selected to not include any strong broad emission from the conical outflow. The \oiii~velocity field is modelled by fitting a two dimensional arc-tangent disc model given by

\begin{equation}
    V(r)=\frac{2}{\pi}V_{max}\arctan \Big( \frac{r}{r_{dyn}} \Big),
\end{equation}

\noindent where V(r) is rotation velocity at radius r from the dynamical center, $V_{max}$ is plateau velocity, and $r_{dyn}$ is the radius at which the arc-tangent function has a turn over to a decreasing slope. The measured line-of-sight velocity from our observations relates to V(r) as

\begin{equation}
    V=V_{0} + \sin i\cos\theta V(r),
\end{equation}

\noindent where

\begin{equation}
    \cos\theta = \frac{(\sin\phi(x_{0}-x))+(\cos\phi(y_{0}-y))}{r}.
\end{equation}

\noindent Radial distance from the dynamical center to each spaxel is given by
\begin{equation}
    r = \sqrt{(x-x_{0})^{2}+\Big( \frac{y-y_{0}}{\cos i} \Big)^2},
\end{equation}

\noindent where $x_{0},y_{0}$ is spaxel location of the dynamical center, $V_{0}$ is velocity offset at the dynamical center relative to the redshift of the quasar broad line region, $\phi$ is position angle in spaxel space, and $i$ is the inclination of the disc. 

\begin{figure}[!th]
    \center
    \includegraphics[width=3.7in]{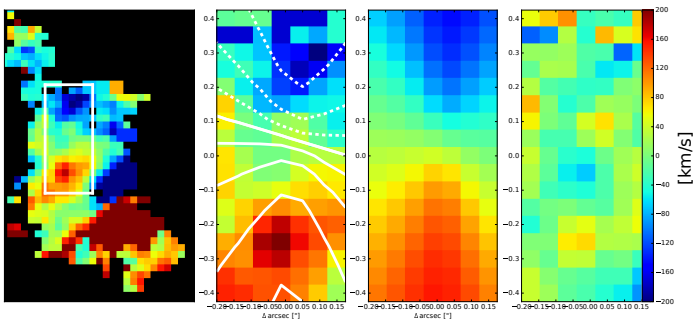}
    \caption{Observed and modeled galactic rotation disc, detected in the nebular \oiii~line in the 3C 298 system. (LEFT) The entire \oiii~radial velocity map with overlaid region used for the modeled disc fit. (MIDDLE LEFT) Observed \oiii~ velocities with the spider diagram overlaid from the disc model. (MIDDLE RIGHT) The best fit 2D disc model. (RIGHT) The residual map shows the observed velocities subtracted by the disc model fit. The dynamical center is at 0,0.}
    \label{fig:rotation_model}
\end{figure}

We fit the seven parameter velocity field model to a selected region of our observed \oiii~velocity map using a non-linear least squares routine. The selected box region tries to exclude broad emission from the extended quasar outflow and a tidal feature that seems to show its own distinct kinematic structure. Our best fit has a $\chi_{R}^{2}=$0.7 with the following values: $x_{0},y_{0} = $ -0.49\arcsec, +0.94\arcsec~relative to the centroid of the quasar, $V_{max}$= 209$\pm33$ \kms, $r_{dyn}=$ 2.2 $\pm0.98$ kpc, $i= 72\pm4^\circ$, $\phi=-109\pm6^\circ$, and $V_{0}=170\pm9$ \kms~relative to the redshift of the quasar broad line region. Figure \ref{fig:rotation_model} shows the region selected for disc fitting with the modeled 2D velocity profile and residuals between the model and observed velocities. The disc is offset from the quasar by 1.06 \arcsec~or $\sim$ 9 kpc, which is evidence of a merging disc system in 3C 298. We obtain a dynamical mass of $1.3\pm0.8\times10^{10}$\msun~within a radius of $r_{dyn}=2.2$ kpc. The dynamical center of this disc is in the vicinity of the isolated AGN outflow identified in Figure \ref{fig:all_spec}. With currently available archival Chandra ACIS observations we are unable to explore the X-ray properties of this secondary AGN candidate. Unfortunately there is strong asymmetry in the Chandra PSF along this position angle from the quasar. Future Chandra observations at a specific spacecraft roll angle will allow us to search for X-ray emission from this secondary AGN and explore the potential dual AGN nature of this system.

\subsection{ALMA: Data reduction}
Data reduction was performed using CASA (Common Astronomy Software Applications \citep{McMullin07}) version 4.4 and 4.7. At the observed frequencies, synchrotron emission from the quasar dominates over molecular CO or dust emission, and has sufficient SNR to perform self calibrations directly on the science source. We used the CASA \textit{CLEAN} function to establish a model for the synchrotron continuum through several interactive runs with clean masks centered on high SNR features. Cleaning is performed with Briggs weighting using a robust value of 0.5 with a pixel scale of 0.05\arcsec. We used the \textit{gaincal} function to perform phase corrections. The self-calibrated data was then cleaned again with further phase corrections until we did not see a significant improvement in the SNR on the continuum. The final root mean square (rms) improved by a factor of 6-8 in the continuum images. We perform continuum subtraction in UV-space by fitting a first order polynomial to channels free of line emission from the host galaxy, and subtract this fit from the rest of the channels.

Data cubes were imaged using clean with a plate scale of 0.05\arcsec~per pixel, spectral channel size of 34 \kms~and natural weighting to improve SNR in diffuse structure of the host galaxy at cost of angular resolution. For cycle 3 data, clean masks were placed on CO emission with SNR$>$5 with the same mask applied to all channels. The resulting beam sizes and rms values are reported in Table \ref{ALMAtable:summary}.

We find the cycle 2 band 4 data cube is a factor of 2 noisier due to poorer weather conditions at the time of observations. Faint emission is detected with a peak SNR of 3.6$\sigma$ for CO (J=3-2) emission over a number of spaxels in the east/north-east direction from the quasar. Applying UV tapering to a resolution of 0.98\arcsec$\times$0.8\arcsec with Briggs weighting (robust = 0.5) improves the SNR to 5 however the line is not spatially resolved. Therefore for the analysis of CO (J=3-2) we defer to cycle 3 data taken under superior weather conditions.  

We align the ALMA and OSIRIS data cubes by matching the position of the optical quasar emission to the unresolved quasar synchrotron emission centroid in the continuum map. The centroid of the quasar matches the location of the unresolved point source with the flattest spectra seen using high resolution cm observation of 3C 298 with MERLIN (component B1/B2 in \citep{fanti02}). This location is typically associated with the quasar core. Finally, we rotate the ALMA data cube to a position angle of 103$^{\circ}$ to match the OSIRIS observations.

\subsection{ALMA: Analysis}\label{sec:ALMA_analysis}

\begin{figure*}[!th]
    \center
    \includegraphics[width=7.8in]{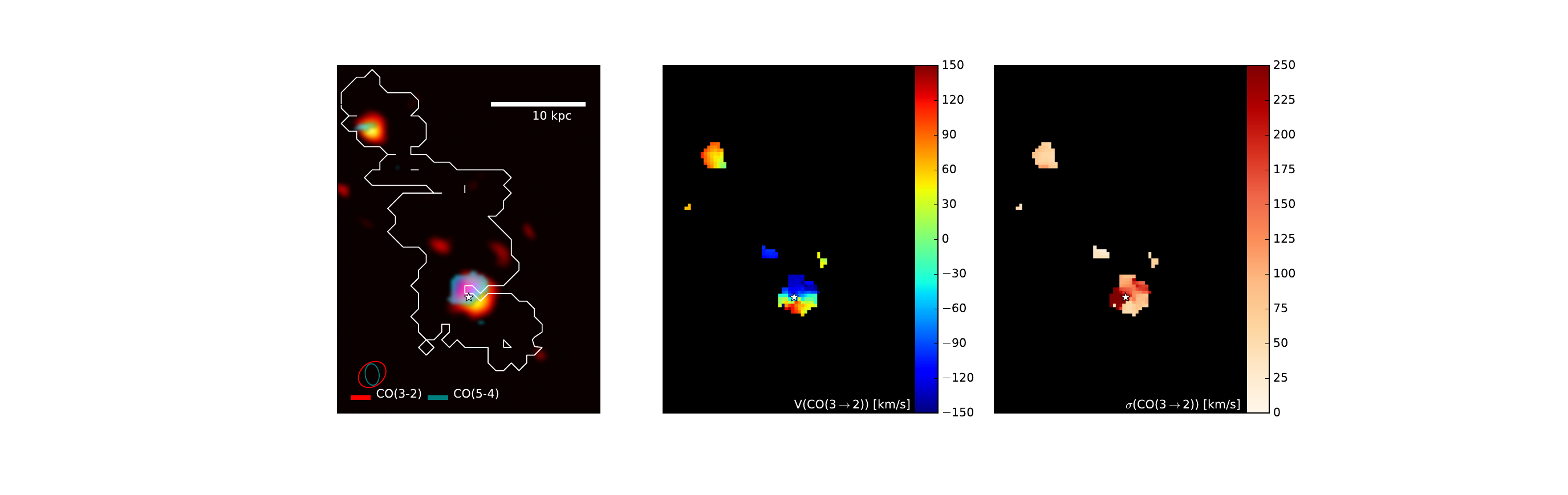}
    \caption{ALMA band 4 and 6 CO (J=3-2) and (J=5-4) observations of 3C 298. (LEFT) Integrated emission shown in red for CO (3-2) and teal for CO (5-4) with white contours representing the lowest surface brightness of the integrated \ha~map. (MIDDLE) Radial velocity map (\kms) for CO (3-2) data relative to the redshift of the quasar. (RIGHT) Radial velocity dispersion map (\kms) of CO (3-2) emission. We find that majority of the molecular gas resides in a disc centered on the quasar and in a molecular clump 16 kpc away from the quasar where active star formation is present as indicated by the \logohb~and \loghn~line ratios (blue region in Figure \ref{fig:n2ha_BPT}). The synthesized beam is shown in the lower left corner. White star marks the location of the quasar}
    \label{fig:alma}
\end{figure*}

In this section we describe the analysis of resolved CO (3-2) and (5-4) 3D spectroscopy. Initial inspection revealed high SNR $\gtrsim5$ detection of CO (J=3-2) and (J=5-4) emission. We extracted spectra over the beam size centered on the detected regions. We then collapsed the data cube over spectral channels where the CO features are detected, and constructed SNR maps by dividing the integrated emission by the standard deviation computed in an empty sky region. Spaxels that showed an emission line with a SNR $\geq$ 3 are fit with a Gaussian profile. We constructed a flux map by integrating over the line ($-3\sigma$ to $+3\sigma$). We created a velocity map by computing the Doppler shift of the emission line centroid relative to the redshift (z=1.439) of the broad line region (calculated from \hb), and generated a velocity dispersion map from the Gaussian fit. Figure \ref{fig:alma} shows the CO emission, radial velocity, and dispersion maps. 

\subsubsection{Molecular gas velocity field modeling}
\begin{figure*}[!th]
    \center
    \includegraphics[width=7.3in]{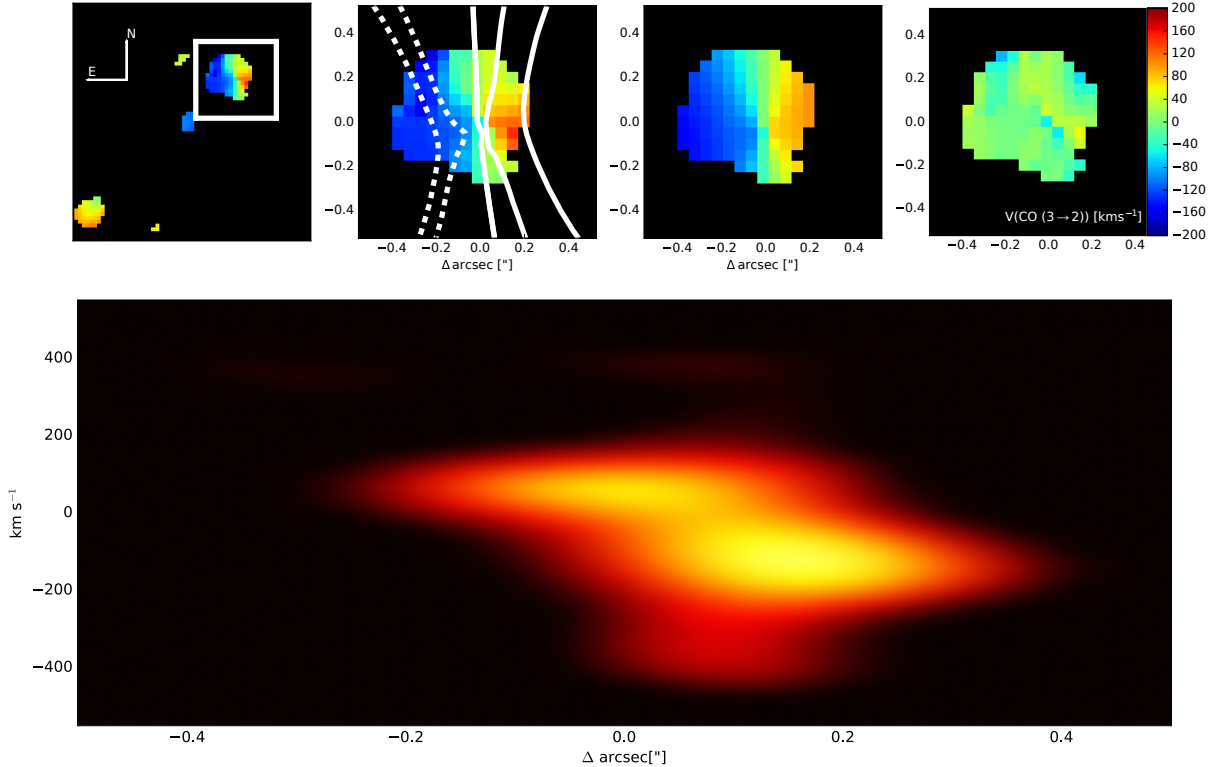}
    \caption{(TOP LEFT) Velocity field modeling of the molecular disc. CO (3-2) radial velocity map with a box highlighting the location of the molecular disc. (MIDDLE LEFT) Boxed region with spider diagram of the disc model. (MIDDLE RIGHT) The best fit rotating disc model. (RIGHT) The residuals between the observed velocity and model fit. The quasar is located at 0,0. The fitting was done on data at the ALMA observation position angle of 0$^{\circ}$ to avoid any pixel interpolation that may cause a difference in the fitting results. (BOTTOM) Position velocity (PV) diagram along the major axis of the ALMA disc, which illustrates the the velocity field and plateau velocity on kiloparsec scale. Broad blue shifted emission that is not associated with the disc's rotation is observed in the PV diagram from -400 to 200 \kms as the large molecular outflow.}
    \label{fig:co_vel_model}
\end{figure*}

In this section we outline modeling performed on the CO (3-2) radial velocity map to derive a dynamical mass for the inner $\sim$2 kpc of the host galaxy. We find the majority of the CO emission to be concentrated at the spatial location of the quasar and with a secondary peak offset by 16 kpc spatially coincident with the recent star formation from OSIRIS BPT analysis (see Figure \ref{fig:n2ha_BPT} and section \ref{sec:OSIRIS_BPT}). CO (3-2) emission concentrated on the quasar shows a distinct velocity gradient resembling a rotating disc, with an extent of about 0.8\arcsec~with a velocity difference of $\pm$150 \kms. Similar to section \ref{sec:vel_model} we model the molecular disc with a hyperbolic arctangent function and derive the following properties about the disc: $\chi_{R}^{2}=$0.6: $x_{0},y_{0} = $ 0.05\arcsec, 0.0\arcsec~relative to the centroid of the quasar, $V_{max}$= 392$\pm65$ \kms, $r_{dyn}=$ 2.1 $\pm0.9$ kpc, $i=54.37\pm6.4^\circ$, $\phi=5.3\pm1.28^\circ$, and $V_{0}=-13.0\pm3.15$\kms. Fit of the disc along with the residuals are shown in Figure \ref{fig:co_vel_model}. We derive a dynamical mass 1.7$\pm0.9\times10^{10}$\msun within $r_{dyn}=$2.1 kpc using the disc inclination derived from the velocity modeling. 

Alternatively, we collapsed the data cube along the spectral direction from -150 \kms~to +150 \kms~to produce a CO (3-2) image of the molecular disk. Using a 2D Gaussian we obtain a major axis value of 0.374$\pm0.05$\arcsec and a minor axis of 0.155$\pm$0.098\arcsec~corrected for the beam size. Assuming the molecular disc can be approximated by an oblate spheroid we derive the inclination angle using the following formula \citep{Holmberg46}:

\begin{equation}
    \rm i = cos^{-1}\sqrt{\frac{(b/a)^2-q_{0}^{2}}{1-q_{0}^{2}}},
\end{equation}

\noindent where a and b are the major and minor axes, and $\rm q_{0}$ is the axial ratio for an edge on disc taken to be 0.13. We derive a dynamical mass of $1.0\pm0.5\times10^{10}$ \msun~using a major axis radius of 1.6$\pm0.215$ kpc and an inclination angle of $66.6\pm17.8^{\circ}$. The dynamical masses and disc inclination angles obtained with these two methods agree within the error. With two distinct galactic discs (see section \ref{sec:vel_model}) in the 3C 298 system we are finding evidence for an intermediate-late stage merger.

\cite{Shen11} determined a single epoch black hole mass for 3C 298 using three calibration methods: $10^{9.57\pm0.03}$ \msun~using the MgII-black hole mass relationship calibrated from their study; $10^{9.37\pm0.03}$ \msun~using the \cite{VO09} calibration; and $10^{9.56\pm0.01}$ \msun~using \cite{MD04} relation. The systematic differences between black hole mass measurements are typically due to differing quasar samples, and properties of the emission lines (e.g., \hb) used for the SMBH mass calculation. 

\begin{figure}
    \includegraphics[width=3.2in]{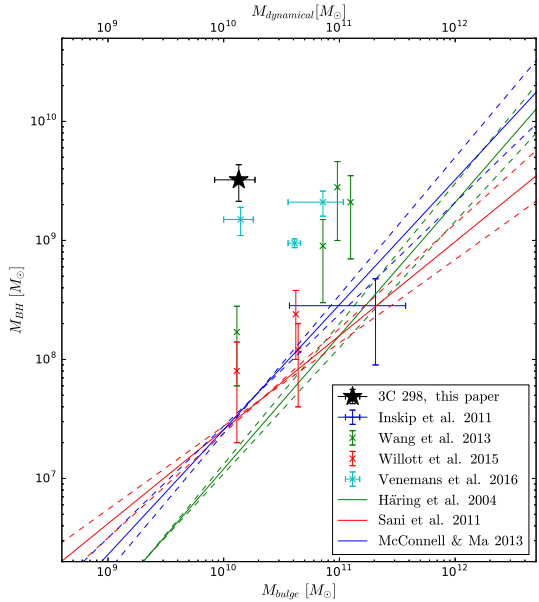}
    \caption{Local SMBH vs galactic bulge dynamical mass relationship compared to 3C 298 measured values. The black star represents the average dynamical mass derived from the intensity and kinematic molecular gas disc modeling and the average black hole mass derived with \cite{MD04,VO09,Shen11} relationships. Errors are added in quadrature from the various methods used to derive the dynamical mass of the disc and the black hole mass. Local scaling laws from various studies \citep{Haring04,Sani11,McConnell13} are shown in solid lines with their respective 1$\sigma$ errors shown with dotted curves. The observed black hole mass and dynamical mass of the bulge of 3C 298 resides $\sim$2-2.5 orders of magnitude away from the local scaling relation. Points from literature \citep{Wang13,Willott15,Venemans16} for z$>5.7$ quasars are also included. There is line of evidence that some of these high redshifts systems also reside off the local scaling laws.}
    \label{fig:mbhmb}
\end{figure}

With both the derived dynamical and black hole mass we can compare the nuclear region of 3C 298 to the local black hole M$\rm_{bulge}-M_{BH}$ relationship. In Figure \ref{fig:mbhmb} we plot the 3C 298 black hole and bulge dynamical mass relative to the local scaling relation from \cite{Haring04,Sani11,McConnell13}. Compared to all three relationships the 3C 298 bulge dynamical mass is either $\sim$2-2.5 orders of magnitude lower than what is expected or the black hole is $\sim$ 2 orders of magnitude higher than what is expected for its bulge mass. If one assumes the 3C 298 bulge is more extended than what is measured by ALMA, then we can use the tidal feature at 21.5 kpc as an estimate for the total enclosed mass. To do this, we integrate all spaxels with a detected CO (3-2) line and remove the outflow contribution by simultaneously fitting a broad and a narrow Gaussian component to the outflow region. We measure a maximum line width of 369.3 \kms~at 90$\%$ intensity, and find the maximum circular velocity of 184.6/$\sin(i)$ \kms at 21.5 kpc from the quasar. Using this maximum circular velocity, we obtain a total enclosed mass at this radius of $\sim2.7\times10^{11}$ \msun~assuming the smallest measured inclination angle from the molecular disc. This value should be considered as a maximum limit for the total enclosed mass of the 3C 298 system. This enclosed dynamical mass at 21.5 kpc from the quasar still places 3C 298 an order of magnitude above from the local scaling relation. Also, the radius used for this calculation is significantly larger than the typical bulge radius at this redshift \citep{Bruce14,Sachdeva17}, and there is no guarantee that the entire enclosed stellar mass will end up in the newly created bulge once the merger is finished. This implies a delay in growth of the galactic bulge for the 3C 298 host galaxy with respect to its SMBH. Averaging SMBH masses and dynamical masses determined for 3C 298 we obtain the following values: $\rm M_{dyn,bulge}= 1.35\pm0.5\times10^{10}$, $\rm M_{SMBH}=3.23\pm1.1\times10^{9}$. Errors are added in quadrature with the SMBH mass standard deviation included to address the systematic uncertainty associated with using various calibration methods.

\subsubsection{Molecular gas mass}
In this section we derive the molecular gas mass from spatially integrated CO (3-2) and (5-4) emission presented in Table \ref{tab:flux_CO}.

\begin{deluxetable}{lcc}

\tablecaption{Measured molecular line intensities\label{tab:flux_CO}}
\tablehead{\colhead{Region} &\colhead{$\rm I_{CO (3-2)}$} & \colhead{$\rm I_{CO (5-4)}$} } 
\startdata
Molecular disc & 0.63$\pm$0.035 Jy \kms & 1.26$\pm$0.063 Jy \kms\\
Tidal Feature\tablenotemark{a} & 0.13$\pm$0.013 Jy \kms& 0.1$\pm$0.025 Jy \kms\\
\enddata
\tablenotetext{a}{Molecular gas associated with star forming region labeled in blue in Figure \ref{fig:n2ha_BPT}}
\end{deluxetable}

In order to measure the molecular gas in the 3C 298 system we convert the CO (3-2) flux into $L^{'}_{CO(3-2)}$ using the following equation from \cite{CarillinWalter13},

\begin{equation}
   \rm L^{'}_{CO(3-2)}=3.25 \times 10^{7}S_{CO(3-2)}\Delta v \frac{D^{2}_{L}}{(1+z)^{3}\nu_{obs}^{2}}K~km~s^{-1}~pc^{2}.  
\end{equation} 

\noindent We convert to CO (1-0) luminosity (L$^{'}_{CO(1-0)}$) using a ratio $L^{'}_{\rm CO(3-2)}/L^{'}_{\rm CO(1-0)}$ of 0.97 \citep{CarillinWalter13}. Using the typical $\alpha_{\rm CO}$ value from nuclear star bursts and quasars (0.8 K \kms$\rm pc^{2}$)$^{-1}$), we derive a total molecular gas mass equal to $8.75\pm0.4\times10^{9}$\msun. The total molecular gas includes all regions with a fitted CO (3-2) line in Figure \ref{fig:alma}. The molecular gas disc contains 6.6$\pm0.36\times10^{9}$\msun, while the active star formation region that resides 16 kpc away contains 1.44$\pm0.14\times10^{9}$\msun. We measure a 2$\sigma$ molecular gas limit of $\rm 1\times10^{9}$\msun$\rm(\frac{\alpha_{CO}}{0.8})(\frac{V_{FWHM}}{200 km s^{-1}})$ per beam. This corresponds to a molecular gas surface density of 104 \msun $\rm pc^{-2}(\frac{\alpha_{CO}}{0.8})(\frac{V_{FWHM}}{200 km s^{-1}})$. These estimates are valid for the inner few arcseconds around the phase center (quasar). 

\begin{figure*}[!th]
    \center
    \includegraphics[width=7.8in]{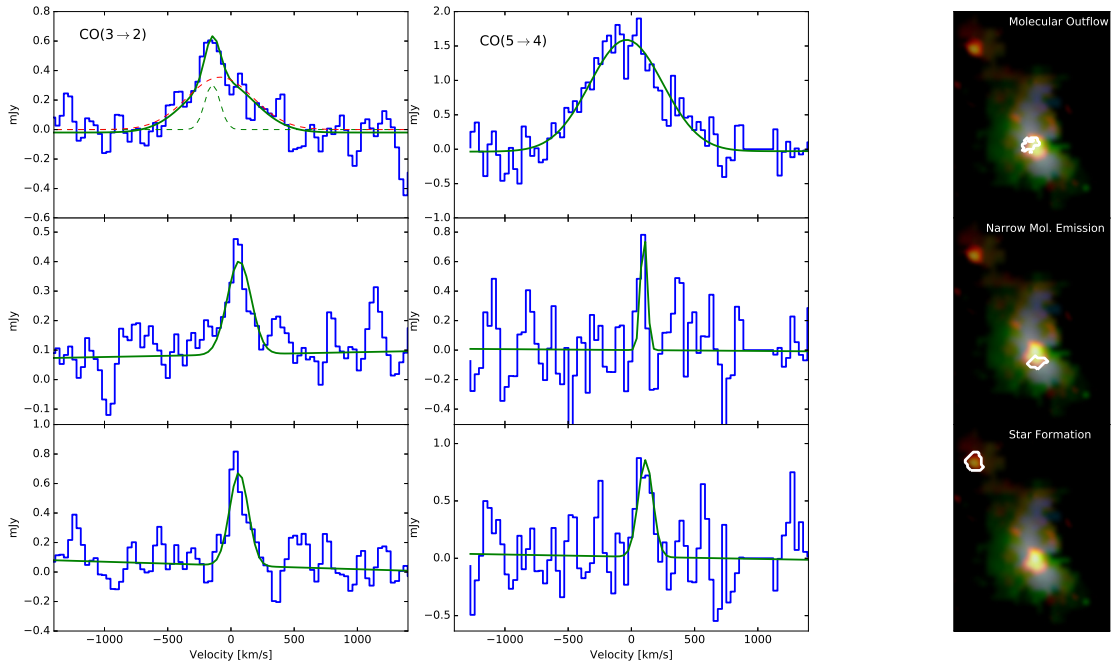}
    \caption{Spectra of molecular CO (3-2) and (5-4) emission in distinct regions of the host galaxy. Left and middle columns show CO (3-2) and (5-4) emission spectra respectively, while the right column showcases a three colour composite from nebular emission lines \oiii, \ha, \sii~with an overlay of CO (3-2) emission in orange. White contours represent the region over which the cubes were spatially integrated to construct the spectra. (TOP) Spectrum constructed by integrating over region with broad ($\rm V_{\sigma}>270$\kms) emission implying a molecular outflow. Both lines are fit with a broad Gaussian profile, CO (3-2) requires a second narrow component that is likely emission from the disc. (MIDDLE) Spectrum constructed over region with narrow emission in the molecular disc, both spectra are well fit with a single Gaussian component. (BOTTOM) Integrated spectrum over the molecular reservoir in the star forming region identified in Figure \ref{fig:n2ha_BPT}. Both CO (3-2) and (5-4) spectra are fit with a narrow component.}
    \label{fig:alma_spec}
\end{figure*}

CO (3-2) and (5-4) emission in some spaxels shows relatively broad ($V_{\sigma}>270$\kms) emission, greater than the escape velocity $\sqrt{2}V_{rotational}\sim$270 \kms~at the observed edge of the rotating disc. In the position-velocity diagram (Figure \ref{fig:co_vel_model}) broad blue-shifted emission is seen with velocities ranging from -400 to 200 \kms~that spatially resides away from the ordered $\pm$150\kms~rotation profile. Integrating over these spaxels the CO (3-2) and (5-4) emission profiles resemble both an outflow with broad emission, and narrow emission that is likely emanating from the molecular disc. We fit a double Gaussian component to the CO (3-2) line and a single Gaussian to the CO (5-4) data. We measure FWHM of 624$\pm$49 \kms~for the broad line emission in CO (3-2) and 687$\pm$18 \kms in CO (5-4). The broad emission line represents either an extended outflow originating from the nuclear region of 3C 298 quasar or from the molecular disc itself. Figure \ref{fig:alma_spec} shows spectra of the CO (3-2) and (5-4) emission regions with their corresponding Gaussian fits. The broad emission in both lines is predominantly found on the blue shifted side of the rotating disc. The total molecular gas in the outflow is 3.3$\pm0.1\times10^{9}$ \msun. In section \ref{sec:outflow} we measure the molecular gas outflow rate, kinetic energy, and momentum to understand its impact on the galaxy.

\subsection{HST WFC3: PSF Subtraction}

We obtain archival Hubble Space Telescope (HST) observations of 3C 298 to quantify properties of the stellar populations of the host galaxy (GO13023, P.I. Chiaberge). Detailed description of the observations are available in \citep{Hilbert16}. In summary, observations were taken in the F606W ($\lambda_{p}$=588.7 nm, width = 218.2 nm) and F140W ($\lambda_{p}$=1392.3 nm, width=384.0 nm) filters which cover the rest frame wavelength range of 196.6 nm - 286.1 nm and 492.1-649.6 nm, respectively. These wavelength ranges bracket the 4000\AA~break feature in galaxy spectra. Two observations were taken in each filter for a total exposure time of 1100.0s (F606W) and 498.46s (F140W). 

We utilize the nearby star SDSS J141908.18+062834.7 to construct a PSF for the F606W filter. The PSF star and quasar have a similar magnitude in the F606W filter with a similar g-r colour as measured in SDSS. Both the PSF star and quasar are saturated, so they share similar bleeding and diffraction patterns on the detector. In the F140W filter the quasar is unsaturated and is about 0.8 magnitudes brighter, so we combine two unsaturated stars in the field to produce a final PSF with matching SNR in the diffraction spikes structure. We extract a 12\arcsec$\times$12\arcsec~box centered on the quasar and the PSF, and scale the flux of the PSF image to match the peak of the quasar emission and then subtract the two images. We also tried to match only the flux in the diffraction spikes that do not overlap with the structure in the host galaxy, and obtained a similar scaling factor. In both filters the inner 1\arcsec~is dominated by noise from the PSF subtraction, the diffraction spikes at position angles of 0$^{\circ}$, 180$^{\circ}$, 225$^{\circ}$ and along the bleeding pattern in the F606W filter that extends about 0.5" along PA 0$^{\circ}$ from the quasar. However, the majority of the host galaxy lies between position angles of 45$^{\circ}$ and 170$^{\circ}$ where the structure is least affected by residual noise. The quoted residual structure are for an image at a PA of 103$^{\circ}$ matching the observations of OSIRIS and ALMA. 

We convert electron counts at each pixel into flux density (Jy) using the 'PHOTFNU' header value. We resize the pixels to 100 milliarcsecond plate scale to match OSIRIS and ALMA data by using the flux conserving IDL \textit{frebin} routine. We convert the maps into AB magnitude/arcsec$^{2}$ and construct a colour map of the host galaxy. Reliable colours are extracted down to a surface brightness limit of 23.5 AB mag/arcsecond$^{2}$ for the majority of the host galaxy, with the exception of the redshifted outflow region that falls in the area dominated by residual noise from PSF subtraction.

\begin{figure*}[!th]
    \centering
    \includegraphics[width=6in]{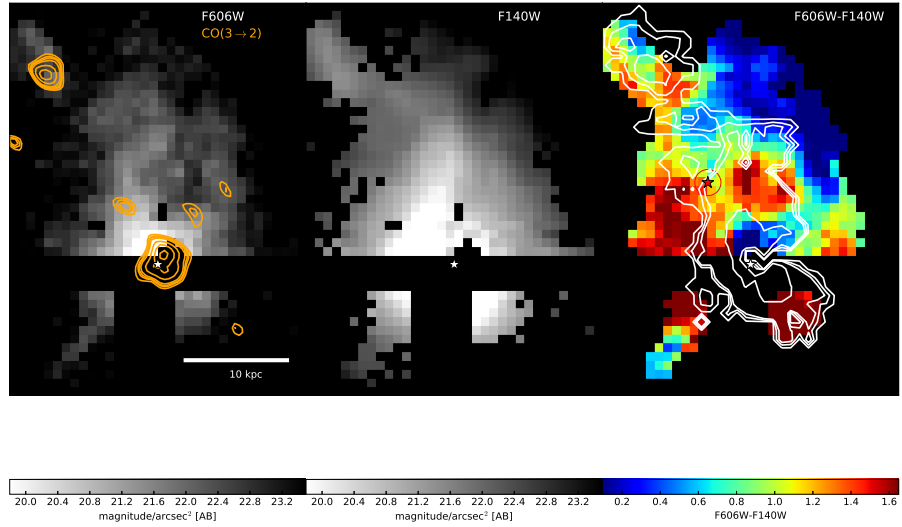}
    \caption{HST WFC3 Rest-frame near-UV (F606W) and approximately rest frame V band (F140W) images of the stellar light in the host galaxy of 3C 298. The images have had the bright unresolved quasar emission subtracted from them. (LEFT) Rest frame near-UV image representing light from young massive stars in the 3C 298 system. Orange contours represent the CO 3-2 ALMA observations of the molecular gas. The UV morphology appears clumpy as in \ha, indicating regions heated by young stars. (MIDDLE) Rest frame V band image which traces light from a combination of young and older stellar populations. The units are in AB magnitude/arcsec$^{2}$. (RIGHT) Colours (F606W-F140W) of stars in the host galaxy of 3C 298 representing the ages of the stellar populations. White contours represent integrated \ha~luminosity tracing the ionized gas; note the similarity between the morphology of the ionized emission and the stellar light. Bluer regions contain younger stars, and are typically found where clumpy structure exists in \ha, CO and UV light; these regions are sights of active star formation. Red star represents the location of the dynamical center of the second merging galaxy in the system.}
    \label{fig:HST}
\end{figure*}

Figure \ref{fig:HST} presents the WFC3 F606W (left) and F140W (middle) after PSF subtraction in 0.1\arcsec/pixel scale. The right panel shows the resolved host galaxy colours (F606W-F140W). Overlaying \ha~contours on the colour map shows that the star forming regions identified in OSIRIS (blue region in Figure \ref{fig:n2ha_BPT}) nicely aligns to regions with bluer colours. Additionally we overlay the ALMA CO (3-2) observation. The molecular clump offset 16 kpc from the quasar matches with bluer regions where clumpy structure is seen in the F606W observations, yielding evidence for young stellar populations over these regions as would be expected from on-going star formation. 

\subsection{HST WFC3: Stellar population}

Two available filters for the 3C 298 system are ideally placed to bracket the break feature at rest-frame wavelength of 4000\AA~found in galaxy spectra. Thus we can easily identify young ($<$100 Myr) stellar populations without any ambiguity, as we attribute blue colours (F606W-F140W $<$1.5) in these filters to stellar populations dominated by young O and B stars. Redder colours in these filters can come from older stellar populations (100 Myr - 1 Gyr) or from young dusty star forming regions, and thus we are unable to say much about older stellar populations without additional filters. 

HST colours provide strong constraints on the resolved star formation and stellar population history across the 3C 298 system. We approximate the HST colour map with the flexible stellar population synthesis (FSPS) code \citep{Conroy09,Conroy10}, using a five parameter star formation history model. We start the FSPS code with an exponential star formation history that is followed by a burst of star formation 2 Gyr later, using a solar metallicity and Salpeter IMF. Using the FSPS code, we find that varying the star formation history and e-folding time scale $\tau$ from 0.1 to 10$^{2}$ prior to the starburst does not significantly affect the age of the stellar population post starburst.

The average F606W-F140W colour over the blueshifted outflow region is best explained with a stellar population that has an age of 25 Myr after the burst. According to ionized gas emission, dust does not affect the age estimated in this region since the \ha~and \hb~ratio suggests a low E(B-V)$\sim$0 value. However ALMA observations reveal that part of the molecular disc centered on the quasar has extened emission into the blueshifted outflow region suggesting that a part of the region can have higher dust extinction. The age of the outflow region and the modelled rest frame U-V colour imply that the stellar population resides in the transition zone between the red sequence and blue cloud in the U-V vs M$_{V}$ colour diagram \citep{Hopkins08,Sanchez04,Bell04}. Furthermore u-r modelled colours at this age suggest that the region lies where post starburst galaxies reside in the u-r vs M$_{stellar}$ diagram \citep{Wong12}. The extracted 3C 298 model spectra shows strong Balmer absorption lines with a strong 4000\AA~break, characteristic of a post-starburst galaxy. Stellar colours over the blueshifted outflow region are consistent with a strong episode of star formation that was abruptly halted. 

In the region where star formation is identified in Figure \ref{fig:n2ha_BPT} based on BPT diagnostics and in the molecular gas clump found in CO, using the same initial conditions in the FSPS code with the outflow region, we find a stellar population with an age of 20 Myr after a burst of star formation. Without any dust correction this age should be taken as an upper limit. Ahead of the jet/blueshifted outflow we find an even younger stellar population. From the \ha~emission we can place a limit of 0.3 \myrkpc~across this region. ALMA observation of the CO (3-2) transition yields a limit of 1$\times10^{9}$\msun~on the molecular gas. This suggests that the current star formation rate is relatively low. However the blue colours indicate recent star formation activity. The stellar populations in this region would then be associated with a young age of 7 Myr after the burst. It is likely that the star formation event was very recent and short (6-10 Myr) due to lack of \ha~emission that would otherwise be present if star formation persisted for a time longer than 6 Myr.

Ages quoted in this analysis could be altered by additional quasar emission from light scattered off dust grains or electrons in the ISM making the colours systemically bluer. However polarimetric observations of 3C 298 reveal that the source is not heavily polarized, with an optical polarization of only 0.77$\pm$0.39$\%$ \citep{Stockman84}. Given that the bluest regions also show star formation from other indicators (e.g, \ha~and CO (3-2) emission), this implies that the majority of UV emission comes from young massive stars and not from scattered quasar light. The young stellar population observed $>$16 kpc beyond the blueshifted outflow is even further away than the region photoionized by the quasar, indicating that ionizing radiation from the quasar does not reach this region.

\section{Dynamics and energetics of outflow regions}\label{sec:outflow}
In this section we explore the dynamics and energetics of the ionized and molecular outflows identified in sections \ref{sec:OSIRIS_kinematics}, \ref{sec:OSIRIS_BPT}, and \ref{sec:ALMA_analysis}. We explore the energy budget from a broad absorption line (BAL) wind and from the quasar jets to investigate the origin of the galactic scale winds.

\subsection{Outflow rates}

The redshifted and blueshifted outflow regions resemble a cone-like structure. For a conical outflow with constant density ($\overline{\rho}$) we use the outflow rate equation from \citep{Cano-Diaz12},

\begin{equation}
    \dot{M}=vR^{2}\Omega \overline{\rho},
\end{equation}

\noindent where $v$ is the velocity of the material in the outflow assumed to be moving at a constant rate over the cone, $R$ is the radial extent of the cone and $\Omega$ is the opening angle of the cone. For $\overline{\rho}=M/V=\frac{3M}{R^{3}\Omega}$ the outflow rate equation simplifies to

\begin{equation}\label{equation:outflow_rate}
    \dot{M}=3\frac{Mv}{R}.
\end{equation}

\noindent Assuming individual ionizing clouds in the outflow have the same density, we can derive an ionized gas mass from line-integrated \ha~luminosity and average electron density over the outflow using \citep{OsterbrocknFerland06}, 
\begin{equation}
    \rm M_{\rm gas~ionized}=0.98\times10^{9}\bigg(\frac{L_{H\alpha}}{10^{43}~ergs^{-1}} \bigg)\bigg(\frac{n_{e}}{100~cm^{-3}}\bigg)^{-1}.
\end{equation}

\noindent The electron density is derived from the line ratio of the \sii~lines in the blueshifted outflow region through using a Gaussian fit to each component of the doublet. We find a ratio of ([SII] 6717/[SII]6731=1.14), which yields an electron density of 272 cm$^{-3}$ using the \textit{getTempDen} code part of \textit{PyNeb} \citep{Luridiana15} package. We calculate the ionized gas mass from the \oiii~line using methodology presented in \citep{Cano-Diaz12}. This method requires knowledge of the ionized gas-phase metallicity at the site of the outflow region. We assume a metallicity value set to solar for a conservative estimate (i.e., for less than solar the gas mass increases). For the outflow velocity we use the v$_{10}$ parameter, which is the velocity value where 10$\%$ of the line is integrated. This value is calculated from the model fits to the spatially integrated \oiii~and \ha~lines for each outflow region. The wings of the emission lines most likely give the true average velocity of the outflow, as the lower velocities seen in the line profiles are probably due to projection effects of the conical structure \citep{Cano-Diaz12,Greene12}. Measured properties of the the outflow regions are presented in Table \ref{table-outflow}.

For the redshited and blueshifted outflow regions we obtain a combined outflow rate of 467 \myr~and 1515 \myr~for \oiii~ and \ha, respectively, while for the AGN outflow region we obtain an outflow rate of 9 \myr~and 115 \myr~for \oiii~and \ha. The \ha~outflow rate is likely a better representation of the ionized outflow rate, since \ha~is capable of probing denser regions where \oiii~would be collisionally de-excited. The observed \ha~outflow rate is similar to what is observed in other quasar surveys \citep{Carniani15}, where they measure a larger outflow rate in \ha~compared to \oiii. The total kinetic energy and luminosity of the ionized outflow are 3.5$\times10^{58}$ erg and 1.2$\times10^{45}$ \ergs. The kinetic luminosity is consistent with the value usually invoked in theoretical models \citep{Hopkins12} as necessary to affect star forming properties in the host galaxy. Over the outflow regions we measure a 2$\sigma$ star formation rate upper limit of 0.3 \myrkpc, by using a Gaussian profile with 2$\sigma$ peak flux and FWHM of 200 \kms, with the multi-Gaussian fit of the outflowing gas subtracted. The properties of the multiple outflow components are summarized in Table \ref{table:outflow-properties}

Similarly using equation \ref{equation:outflow_rate} we compute the molecular outflow rate. As found in section \ref{sec:ALMA_analysis} the extent of the broad molecular emission from CO (3-2) and (5-4) lines is approximately 1.6kpc, the total molecular gas mass in the outflow region is 3.3$\times10^{9}$\msun~with an outflow velocity of 400\kms as measured by $V_{10}$ parameter. This yields a total molecular outflow rate of 2300\myr. The kinetic energy and luminosity of the molecular outflow are 5.1$\times10^{57}$ erg and 1$\times10^{44}$ erg/s.

\begin{table*}
\center
\begin{threeparttable}
\centering
\caption{3C 298 Ionized Outflow Properties\label{table:outflow-properties}}
\label{table-outflow}
\medskip
\begin{tabular}{l|cccccccc}
\hline
Region & L$_{\rm H\alpha}$&L$_{[\rm OIII]}$& M$_{\rm H\alpha}$&M$_{[\rm OIII]}$ & $v$ & $R$ & $\dot{\rm M_{H\alpha}}$ & $\dot{\rm M_{[OIII]}}$\\
& $\times10^{44}$ \ergs & $\times10^{44}$ \ergs &$\times10^{8}$\msun&$\times10^{8}$\msun & \kms & kpc & \myr & \myr \\
\hline
Outflow-R & 0.29 & 1.8 & 10 & 3.5 & 1703 & 4.7 & 1110 & 390\\
Outflow-B & 0.09 & 0.3 & 3.2 & 0.6 & 1403 & 3.4 & 405 & 77\\
Outflow-AGN & 0.03 & 0.04 & 0.8 & 0.064 & 1400 & 3 & 115 & 9\\
\hline
\end{tabular}

\end{threeparttable}
\end{table*}

\subsection{Virial parameters \& gas pressure}
Using the virial parameter, a ratio of the free fall time scale to the dynamical time scale we can investigate whether the gas is gravitationally bound or unbound in the outflow regions and molecular disc. A value of unity or below would suggest that the gas is bound and is able to collapse to form stars. A value much greater than one would suggest that the gas is unbound and therefore at present time should not be collapsing to form stars. Over the outflow region we measure a virial parameter 
\begin{equation}
    \alpha_{\rm vir}=5{\sigma^2 R\over GM} \approx 164 
    \left(
    {\sigma\over 500\,{\rm km\, s}^{-1}}
    \right)^2,
\end{equation} 
\noindent using an average velocity dispersion of $\sigma =500\,{\rm km\,s}^{-1}$ seen in the ionized outflow, a radius of $R=3$ kpc, and the ionized gas mass (1.32$\times 10^{9}$ \msun). Using the molecular gas mass of 6.6$\times10^9$ \msun~an average velocity dispersion of 270 \kms~seen in the molecular outflow and a radius of 1.6 kpc yields $\alpha_{\rm vir}$ of 20.5. Both of these virial parameter estimates suggest that at the present time, the physical conditions in the ionized and molecular outflow regions and in the molecular disc are stable against gravitational collapse, hindering star formation. In contrast, the molecular gas clump in the star forming region observed in CO (3-2) yields a virial parameter value of $\alpha_{\rm vir}\approx0.7$. A value close to unity suggests that star formation should be able to proceed in this region, where there are no powerful outflows. 

If the quasar-driven wind does not efficiently cool and adiabatically expands, then the initial driving mechanism is capable of transferring its kinetic energy into mechanical energy in the galactic-scale outflow. The momentum flux of the outflow should therefore be greater than $L_{\rm bol}$/c \citep{Zubovas12}. We measure total momentum flux ($\dot{P}_{\rm outflow}=\dot{M}\times v$) of 2.1$\times10^{37}$ dynes over the entire outflow region. We use the 3000\AA~quasar luminosity to compute a bolometric luminosity value of 1.04$\times10^{47}$ \ergs\, \citep{Runnoe12}. Comparing the outflow momentum flux to the quasar radiation momentum flux ($L_{\rm bol}$/c) of 3.5$\times10^{36}$ dynes, yields a ratio $\dot{P}_{\rm outflow}/\dot{P}_{\rm quasar}$ (loading factor) of 6. In principle this ratio is a lower limit, since there can still be diffuse molecular gas in the ionized outflow regions whose CO emission is below the sensitivity of ALMA. Based on theoretical work by \citealt{Faucher12,Zubovas12} the measured loading factor of $>$6 suggests that the wind is energy-conserving. The total kinetic luminosity of the outflow is also about 2$\%$ of $\rm L_{bol}$, consistent with the above theoretical prediction. 3C 298 observed extended outflow properties strongly suggests that radiation pressure by itself is unable to drive the outflow. Star formation is also incapable of driving the observed outflow as the expected terminal velocity for a supernova driven galactic scale wind is only $\sigma\sim$200 \kms \citep{murray05}, far lower than the observed velocities ($\sigma\sim500$\kms) seen in the bi-conical wind.

The outflow could be induced by either the jet and/or broad-absorption line (BAL) winds from the quasar. We infer a jet pressure 

\begin{equation}
    \rm P_{jet}\approx\frac{L_{jet}\times t}{3V},
\end{equation}

 \noindent of $3\times 10^{-8}$ dynes cm$^{-2}$ using the jet kinetic luminosity of 1.4$\times10^{47}$ erg/s, derived using the 1.4 GHz (570 MHz observed) radio flux-jet power relation \citep{Birzan08,Cavagnolo10}, assuming spherical volumes at a radius of 9 kpc and a time scale of 3 Myr. This is an order of magnitude calculation given the approximations in the jet luminosity, lifetime (t) of the jet, and volume. Observations of other BAL quasars indicate that the kinetic luminosity of outflows can get as high as $10^{46}$ \ergs \citep{Chamberlain15}, which is close to the kinetic luminosity for the 3C 298 jet and would exert a pressure of $10^{-8}$ dynes cm$^{-2}$. This means that a BAL-type wind could potentially drive the extended 3C 298 outflow. However, there is no observed BAL in the optical and infrared spectra of 3C 298. The lack of an observed BAL in 3C 298 could be due to an orientation effect where our line-of-sight does not overlap with the absorbing outflowing gas. If this is not the case, and there is no BAL, the jet of 3C 298 still has the necessary ram pressure capable of inducing a large scale outflow. 
 
The gas pressure ($P=nkT$) in the ionized (\sii) gas is $\sim4\times10^{-10}$ dynes cm$^{-2}$. From the ALMA CO (3-2) observations we measure a molecular gas surface density of 490 \msun pc$^{-1}$ indicating a molecular gas pressure ($P=(\pi/2) G \Sigma_{molecular}^2$) of $\sim1\times10^{-9}$ dynes cm$^{-2}$. Both the BAL wind and quasar jet pressures are 2-3 orders of magnitude higher than the current pressure of the ionized and molecular ISM. The fact that the ionized gas pressure is comparable to or smaller than the weight per unit area of the molecular gas shows that the jet is not currently producing an overpressure (which would be reflected in the thermal gas pressure of the ionized gas in the ISM) relative to the overburden of the molecular gas. This suggests that the jet is now venting out of the galaxy. The interaction between the ISM and the jet and/or BAL wind in the past was likely in a denser environment, and thus confined to a smaller volume with a higher pressure, to initially generate the outflows. 

Both a BAL quasar wind and jet have enough kinetic energy to drive galactic scale winds. These large outflows in 3C 298 supply the necessary energy and momentum to hinder star formation along their path and expel large amounts of gas out of the galaxy. If the sources of turbulence were to halt at the present time, the shocks would dissipate on a dynamical time scale of 3-6 Myr. By that time the majority of the gas in the ionized outflow will have already escaped into the intergalactic medium with most of the molecular gas in the disc removed from the inner few kpc and potentially swept up into the ionized outflow. Any left over gas in the galaxy would quickly cool back into a molecular state. Taken together this shows both negative and ejective feedback occurring along the outflow region, thereby impacting the stellar mass growth of the quasar host.

\section{Discussion}\label{sec:discussion}

\begin{figure}[!th]
    \center
    \includegraphics[width=3.5in]{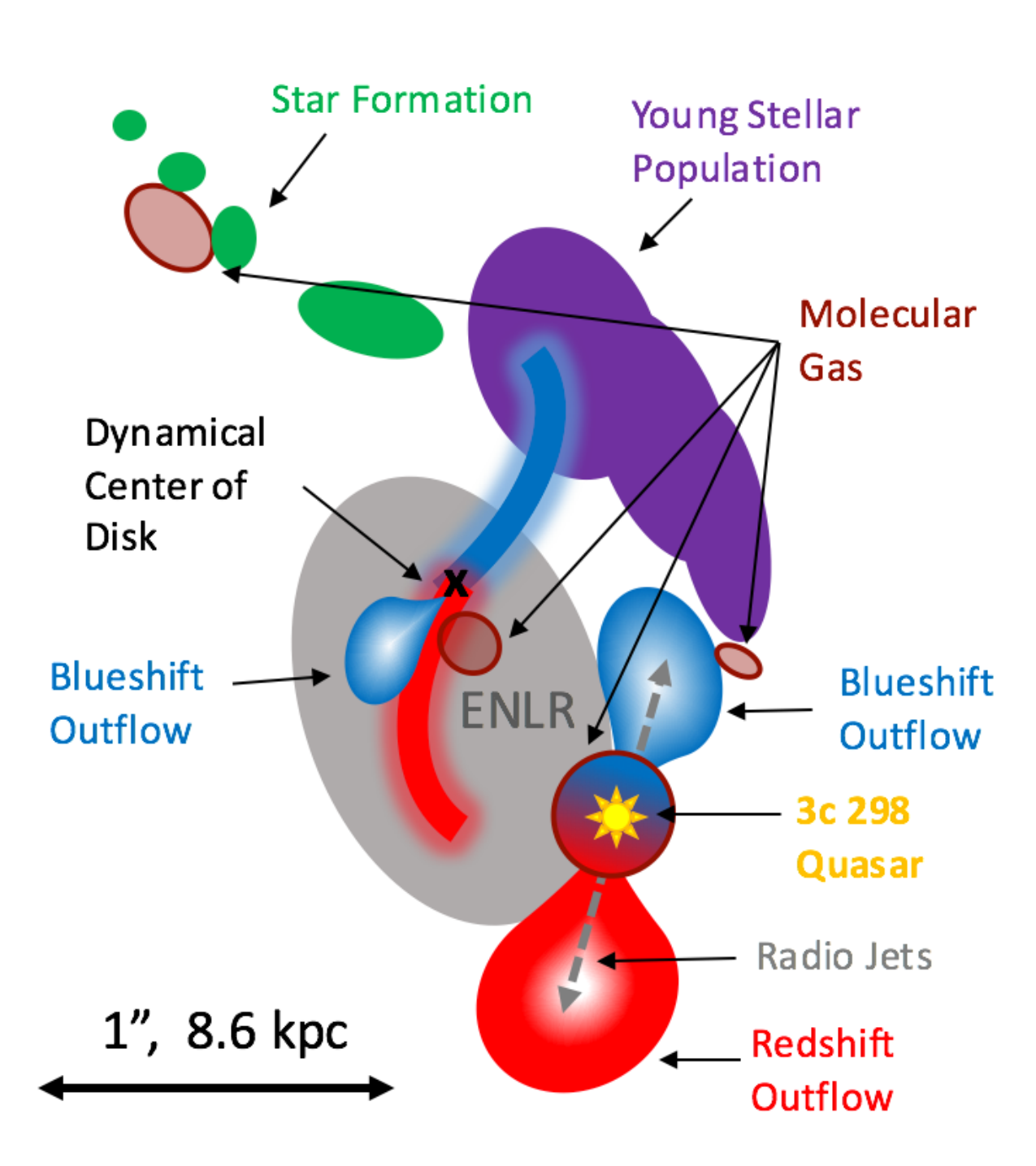}
    \caption{Cartoon illustration of the merger in the 3C 298 system that includes the results from OSIRIS, ALMA, HST, and VLA. The location of the quasar is distinguished by a yellow star. The large blue and red-shifted outflows emanating from the quasar observed in the nebular emission lines are co-aligned with the radio jet (indicated by a grey dashed arrow), where negative feedback is observed. The locations of the ALMA CO (3-2) and (5-4) emission are shown with red-brown ovals, representing molecular gas reservoirs. A molecular disc centered on the quasar is shown with a blue-red gradient. We find strong evidence for molecular outflow on the blueshifted side of the disc. The dynamical center of the secondary merging galaxy is shown with a black 'x' sign. The radial velocity map of this galaxy is well-fit to disc model. A secondary outflow region is found to reside near the dynamical center of the merging galaxy, providing tentative evidence for a secondary AGN in the system. The secondary AGN has spatially-concentrated elevated emission line ratios that are consistent with ionization from an AGN. A large area of the 3C 298 system is observed to have extended narrow line region (ENLR; grey) between both the quasar and putative AGN. The majority of the star formation (green) is occurring in compact clumps offset from both galaxies, which would be consistent with a tidal feature being induced by the merger. A young stellar population (purple) measured from the HST WFC3 colours is offset from both the outflows and molecular gas of the system.}
    \label{fig:cartoon}
\end{figure}

To generate a comprehensive picture of 3C 298, we combine multi-wavelength data to spatially resolve the ionized gas, molecular gas, and stellar light distribution of the host galaxy. This allows a unique comparison between the star formation histories, photo-ionization mechanisms, and gas-phase properties in a high-redshift quasar host galaxy. In Figure \ref{fig:cartoon} we generalize the multi-wavelength analysis of the 3C 298 system in a schematic diagram and summarize its properties in Table \ref{table:summary}.

We make use of observations from VLA to probe the quasar radio jets, WFC3 data from HST to study the stellar contribution, ALMA to measure the resolved molecular gas morphology and dynamics, and Keck LGS-AO to determine the resolved ionized gas properties. These data show clear evidence of a conical outflow in the host galaxy of 3C 298, driven by quasar jets and/or winds, that directly impact the ISM of the host galaxy. VLA images of 3C 298 reveal synchrotron radio jets feeding lobes that extend over $\sim$ 18 kpc from the quasar in the east-west direction (Figure \ref{fig:O3_maps}). Along the radio lobes, OSIRIS kinematic maps reveal broad ($\sigma\sim800$ km/s) emission lines that are offset up to 600 \kms~from the quasar's systemic redshift, indicating an outflow extending over a significant swathe of the galaxy.

We find direct evidence of negative feedback along the path of the jet and outflow regions of the host galaxy. The term ``negative feedback" has been used loosely in the astronomical literature; herein we define negative feedback as injected energy and momentum in the ISM that inhibits the normal thermal cooling and/or decay of turbulence, thereby extending the time scale for star formation to occur. Galactic-scale feedback may also occur when high velocity outflows remove large fractions of the ISM, thereby impacting the star formation and stellar mass history of the host galaxy. The latter might be called ``ejective feedback". Both negative and ejective feedback may occur simultaneously. The 3C 298 ionized outflow region has a bi-conical shape with a primarily blueshifted approaching side and redshifted receding side of the cone (Figures \ref{fig:O3_maps} and \ref{fig:n2ha_BPT}). Emission line ratios (e.g., \nii/\ha~and \oiii/\hb) allow us to distinguish between various photo-ionization modes of 3C 298's ISM, which is illustrated in the nebular diagnostic diagram Figure \ref{fig:n2ha_BPT}. In both of the outflow regions we detect high nebular emission-line ratios that imply a combination of hard photoionizing radiation from the quasar and shock ionization. We find a total ionized mass outflow rate ($\dot{M}$) of 450-1500 \myr~combined for both outflow regions. ALMA observations of molecular gas reveal a disc centered on the quasar. A large fraction of gas in the molecular disc reveals broad emission associated with a molecular outflow with a rate of 2300\myr. Unlike the ionized outflow the molecular outflow emanates in a single direction from the blueshifted side of the molecular disc. The molecular outflow either originates from the 3C 298 quasar nuclear region or from the disc itself. In the rest-frame of the redshifted ionized receding cone, two broad Gaussian functions are required to fit the profiles of the nebular emission lines (Figure \ref{fig:all_spec}). This indicates an approaching and receding component corresponding to the front and back of a hollow cone that suggests expansion of the outflow along our line-of-sight. 

Combining kinematic maps from OSIRIS and ALMA reveals that the ionized outflow gas most likely originates from the molecular disc since the blue/red-shifted outflow matches the blue/red-shifted region of the disc. Furthermore the blueshifted molecular outflow is on the same side as the ionized. We find no narrow ($\sigma$ $<$200 km/s) \ha~emission in either of the outflow regions, suggesting that the majority of the ionized gas is in a turbulent phase. The outflow carries a significant amount of kinetic energy and it is capable of removing the majority of the galaxy's gas on a time scale of 3 Myr. The virial parameter deduced from the ionized and molecular gas over the outflow regions are much greater than one, implying that the ISM is not self-gravitating, and hence will not collapse to form stars. Similarly, the molecular outflow is both stirring the molecular gas and removing it from the inner regions of the galaxy. Ionization occurring over the outflow region is dominated by shocks, which drives intense heating into the ISM that further inhibits star formation. Over the conical outflow region we derive a conservative limit for \ha~emission due to star formation of $<$ 14 \myr (2$\sigma$), or $\sim$0.3 \myrkpc~using the empirical \ha~luminosity-star formation rate \citep{Kennicutt98}. Colours of the stellar light as seen in HST data projected on the blueshifted outflow region are also consistent with a stellar population that had its star formation abruptly halted within the last $\sim$25 Myr (see Figure \ref{fig:HST}).

In the 3C 298 host galaxy we observe star forming regions that are offset from the outflows. The youngest (7 Myr) stellar populations observed with HST (Figure \ref{fig:HST}) imply that a recent burst of star formation has occurred. These regions are found 10-16 kpc east of the quasar. In addition, beyond the extended narrow-line and outflow regions, using OSIRIS and HST we resolve star forming clumps that reside $\sim$10-22 kpc (blue in Figure \ref{fig:n2ha_BPT}) away from the quasar. These clumps have an integrated star formation rate of 88$\pm$9 \myr, based on the observed \ha~luminosity and empirical star formation rate relation \citep{Kennicutt98}. ALMA CO (3-2) and (5-4) data reveal a molecular clump in the vicinity of this region with a total molecular gas mass of 1.4$\times10^{9}$\msun. Using the HST-measured colours, we find a young stellar population with an age of $\sim$20 Myr in these star forming regions. Therefore, combining results from the resolved molecular gas, stellar populations, and ionized gas yields a consistent picture; we find star formation occurring away from large scale outflows, where the quasar winds and/or jets are stirring the ISM and removing gas.

The kinematic map of the ionized gas in 3C 298 shows a systematic velocity gradient across a large portion of the system. We fit this region with a rotating disc model (Figure \ref{fig:rotation_model}). We find evidence for a second rotating disc in the system that is offset by $\sim$8.6 kpc and $\sim$170 \kms from the quasar centroid and central velocity of the broad line region. The dynamical center of the second disc is in the vicinity of a secondary outflow region, suggesting for the presence of a candidate second nucleus. The offset between the dynamical center of the second disc and the secondary outflow is on the order of 3kpc. We note that this is a typical separation that is found between the stellar centroid and AGN location in local late stage mergers \citep{Liu13,Comerford15,Barrows16}. We infer that the 3C 298 system likely has two massive galaxies with distinct rotating discs that are currently experiencing a close passage, where their galactic nuclei and SMBHs have yet to merge.

We calculated the enclosed dynamical mass of the molecular disc at the location of the quasar to be 1-1.7$\times10^{10}$\msun. With the measured SMBH mass ($10^{9.37-9.56}$\msun) 3C 298 resides off the local scaling M$\rm_{bulge}-M_{BH}$ relationship, which indicates that black hole growth must occur earlier than stellar mass assembly. These results are similar to what has been recently observed in nearby ultraluminous infrared galaxies (ULIRGs) in gas-rich merger stages by \citealt{Medling15}, where they find that these ULIRGs predominantly have early black hole growth and have yet to form the bulk of their stellar mass. Similar results have been found for a small number of z$>6$ quasar host galaxies observed with ALMA, where the black hole to galaxy bulge mass ratio is higher by a factor of 3-4 \citep{Wang13,Venemans16} compared to local galaxies. This is counter to previous theoretical simulations that imply star formation and stellar mass build-up should occur during the gas-rich-merger phase, before the SMBH does its predominant growth when AGN feedback is suspected to transpire (i.e., \citealt{DiMatteo05,Hopkins12b, Hopkins16}). 

Models are still unclear about whether distant quasars should reside above, below or even on an extension of the local SMBH-galaxy scaling relations. Hydrodynamical simulations of $\rm z>6$ quasars by \citealt{Barai17} suggest that negative feedback may drive quasars above the local scaling SMBH-galaxy relations. \citealt{Barai17} argue that conical, symmetric outflows can remove gas efficiently from galactic discs, while still allowing gas to accrete on to the SMBH perpendicular to the bi-conical outflow. Furthermore, the same simulations without quasar feedback tend to grow their stellar mass ahead of the SMBH, driving galaxies below the local correlations. \citealt{Alcazar17} also find that the stellar bulges outpace the SMBH in the absence of feedback. \citealt{Alcazar17} include feedback from bursty star formation, which can limit the growth of the SMBH and keeps the system lying below local SMBH-galaxy scaling relation. 

3C 298 at z=1.439 (age 4.47 Gyr) has quasar negative feedback occurring in a conical outflow early in the gas-rich merger-phase, implying that the majority of the nucleus' stellar mass assembly must occur in later phases if it is to reach the scaling relations of today. Since the majority of the current gas supply in the host galaxy of 3C 298 is incapable of forming stars and will soon be removed from the galaxy through the quasar driven winds, perhaps this assembly happens through dry mergers or newly supplied IGM molecular gas. Even if all of the current gas supply in the molecular disc (6.6$\times10^{9}$\msun) turns into stars it will still be insufficient to bring the bulge of the 3C 298 host galaxy onto the local scaling relation. For the 3C 298 system this means that the SMBH mass is formed earlier and more efficiently than the host galaxy's stellar mass. 

The host galaxy of 3C 298 could be a precursor to local galaxies that are significantly offset from the local scaling relation (e.g, NGC 1277). If there is insufficient accretion from the intergalactic medium to fuel future star formation or if 3C 298 resides in a less dense environment than a typical quasar it could fail to form/accrete enough stars to fall on the local scaling laws. A natural way to limit the growth of the stellar bulge in a system with a powerful AGN is to have strong feedback on the ISM through a bi-polar outflow early in the lifetime of the system \citep{Fabian13}.

In summary, the 3C 298 system shows strong evidence of negative feedback from powerful conical outflows that are halting star formation by driving extremely high velocity turbulence in the ISM and by ejecting large amounts of gas. The dynamical time scale of the outflow is $\sim$3-6 Myr, suggesting that quenching of star formation must occur rapidly. In the 3C 298 system, strong feedback has started fairly early in the merger process, well before the final coalescence of the two galactic nuclei. The system shows multiple concurrent phases of the standard merger model \citep{Sanders88,Hopkins08,Alexander12}, with the co-existence of a luminous type-1 quasar, star formation in merger host galaxies, and candidate secondary AGN that is offset from the quasar. This implies that removal of gas and dust surrounding the SMBHs occurred rapidly and early during the merger phase. This early and short onset of quasar feedback compared to total time scale of the merger ($\sim$ 1 Gyr) may also partially explain why it has been difficult for measuring its effect on the star-forming ISM in distant galaxies.

\section{Conclusion}\label{sec:conclusion}
We present integral field spectroscopy of resolved nebular emission lines; \hb, \oiii, \ha, \nii, \sii together with rotational CO (3-2) and (5-4) spatially resolved emission spectroscopy in the host galaxy of the radio loud quasar 3C 298 (z=1.439). These data are supplemented with archival HST-WFC3 and VLA observations of the rest frame UV and optical stellar continuum and jet/lobe synchrotron emission, respectively. 

\begin{itemize}
    \item We find strong evidence for negative feedback in the host galaxy of 3C 298. Dynamics of ionized gas traced through nebular emission lines show a significant quasar-driven outflow encompassing a large area of the host galaxy. We derive an ionized outflow rate of 467-1515\myr.
    \item Co-spatial with the powerful ionized gas outflow VLA imaging shows extended synchrotron emission from the quasar jet/lobes. Radio data suggests that the jet has sufficient energy to drive the outflow. A BAL wind also has the necessary energetics for potentially driving the outflow.
    \item We detect a molecular gas disc centered on the quasar with a total molecular gas of 6.6$\times10^{9}$\msun~and effective radius of 1.6 kpc.
    \item A powerful quasar driven molecular outflow is detected in the molecular disc with an $\rm \dot{M}_{H_{2}}$ outflow rate of 2300 \myr. The molecular gas in the disc will be depleted by the outflow on a time scale of 3 Myr.
    \item Dynamical modeling of the molecular disc reveals that total mass enclosed in the disc is 2-2.5 orders of magnitude below the expected value from local $\rm M_{stellar-bulge} - M_{BH}$ relationship using the measured SMBH mass of the 3C 298 quasar. 
    \item Several kiloparsecs away from the outflow path we find evidence of star forming regions based on BPT diagnostics, strong UV emission from O and B stars, and evidence for a molecular reservoir with a total mass of 1.4$\times10^{9}$\msun.
    \item Disc modeling of the velocity field traced by \oiii~emission shows a second rotating disc with a dynamical center offset from the quasar by 9 kpc. This suggests evidence for a late stage merger where the disc traces the second merging galaxy in the system. The dynamical center of the disc aligns well with a secondary outflow region seen in nebular emission that does not extend from the quasar and does not align with any structure in the radio map. This suggests a candidate secondary AGN in the 3C 298 system.
    \item These observations taken together imply an early onset of negative feedback with a short quenching time compared to time scale of the galactic merger. Feedback is also happening early in the 3C 298 merger process and is occurring well before the assembly of the 3C 298 host galaxy on the local $\rm M_{stellar-bulge} - M_{BH}$ relationship.
\end{itemize}

\begin{deluxetable}{lc}[!th]
\tablecaption{3C 298 Properties \& Results \label{table:summary}}
\tablehead{\colhead{Parameter} & \colhead{Value}
}

\startdata
z$_{\rm quasar}$ & 1.439 \\
RA & 14:19:08.181 \\
DEC &  $+$06:28:34.79\\
$\lambda$L$_{\lambda}$\tablenotemark{a} &2.0$\times10^{46}$\ergs\\
SFR \ha &88$\pm$9\myr\\
SFR IR \tablenotemark{b} & 930$^{+40}_{-40}$\myr\\
M$_{\rm dust}$\tablenotemark{b}&3.8$^{+0.3}_{-0.4}\times10^{8}$\msun\\
M$\rm_{H_{2}}^{Total}$&8.75$\pm0.4$$\times10^{9}$ \msun \\
M$\rm_{H_{2}}^{Mol Disc}$&6.6$\pm0.36$$\times10^{9}$ \msun \\
M$\rm_{H_{2}}^{SF Region}$&1.4$\pm0.14$$\times10^{9}$ \msun \\
$\rm \dot{M}_{[OIII]}$ & 467 \myr\\
$\rm \dot{M}_{H\alpha}$ & 1515 \myr\\
$\rm \dot{M}_{H_{2}}$ & 2300 \myr\\
Outflow $\rm n_{e}$\tablenotemark{c} & 272 cm$^{-3}$\\
$\rm M_{dyn,bulge}$ & $1.35\pm0.5\times10^{10}$\\
$\rm M_{SMBH}$ & $3.23\pm1.1\times10^{9}$\\
\enddata
\tablenotetext{a}{Computed at rest-frame 3000 \AA}
\tablenotetext{b}{Value from \citep{Podigachoski15}}
\tablenotetext{c}{Value based on [SII] line ratio measurment over blueshifted outflow region.}
\end{deluxetable}

\acknowledgments

The authors wish to thank Randy Campbell and Jim Lyke with their assistance at the telescope to acquire the Keck OSIRIS data sets. We also appreciate valuable discussions with Dusan Keres and the constructive comments made by the anonymous referee. The data presented herein were obtained at the W.M. Keck Observatory, which is operated as a scientific partnership among the California Institute of Technology, the University of California and the National Aeronautics and Space Administration. The Observatory was made possible by the generous financial support of the W.M. Keck Foundation. The authors wish to recognize and acknowledge the very significant cultural role and reverence that the summit of Maunakea has always had within the indigenous Hawaiian community. We are most fortunate to have the opportunity to conduct observations from this mountain. This paper makes use of the following ALMA data: ADS/JAO.ALMA[2013.1.01359.S]. ALMA is a partnership of ESO (representing its member states), NSF (USA) and NINS (Japan), together with NRC (Canada), NSC and ASIAA (Taiwan), and KASI (Republic of Korea), in cooperation with the Republic of Chile. Based on observations made with the NASA/ESA Hubble Space Telescope, obtained from the Data Archive (Program GO13023) at the Space Telescope Science Institute, which is operated by the Association of Universities for Research in Astronomy, Inc., under NASA contract NAS 5-26555. A portion of the research was conducted at The Dunlap Institute for Astronomy and Astrophysics that is funded through an endowment established by the David Dunlap family and the University of Toronto. This research has made use of the NASA/IPAC Extragalactic Database (NED) which is operated by the Jet Propulsion Laboratory, California Institute of Technology, under contract with the National Aeronautics and Space Administration.

\facilities{Keck(OSIRIS-LGSAO), ALMA (Band 4 and 6), HST(WFC3), VLA}

\software{Scipy: \citep{scipy},
          CASA: \citep{McMullin07},
          OSIRIS Data Reduction Pipeline:\citep{OSIRIS_DRP},
          Matplotlib \citep{matplotlib}}

\pagebreak

\end{document}